\documentclass[fleqn,usenatbib]{mnras}

\usepackage{newtxtext,newtxmath}

\usepackage[T1]{fontenc}

\DeclareRobustCommand{\VAN}[3]{#2}
\let\VANthebibliography\thebibliography
\def\thebibliography{\DeclareRobustCommand{\VAN}[3]{##3}\VANthebibliography}

\usepackage{graphicx}	
\usepackage{amsmath}	
\usepackage{caption}
\usepackage{subcaption}
\usepackage[table]{xcolor}
\usepackage{multirow}

\newcommand{\kms}{\,km\,s$^{-1}$} 

\title[A. M. Sebastian et al.]{E-XQR-30: The evolution of \ion{Mg}{ii}, \ion{C}{ii} and \ion{O}{i} across $2<z<6$}

\author[A. M. Sebastian et al.]{
Alma Maria Sebastian,$^{1,2}$\thanks{E-mail: asebastian@swin.edu.au}
Emma Ryan-Weber,$^{1,2}$
Rebecca L. Davies,$^{1,2}$
George D. Becker,$^{3}$
\newauthor\ Laura C. Keating,$^{4}$
Valentina D'Odorico,$^{5,6,7}$
Romain A. Meyer,$^{8}$
Sarah E. I. Bosman,$^{9,10}$
Guido Cupani,$^{5,7}$
\newauthor\ Girish Kulkarni,$^{11}$ 
Martin G. Haehnelt,$^{12}$
Samuel Lai,$^{13}$
Anna--Christina~Eilers,$^{14}$
Manuela Bischetti,$^{15}$
\newauthor\ Simona Gallerani $^{6}$
\\
$^{1}$Centre for Astronomy and Astrophysics, Swinburne University of Technology, Hawthorn, Victoria 3122, Australia\\
$^{2}$ARC Centre of Excellence for All Sky Astrophysics in 3 Dimensions (ASTRO3D), Australia\\
$^{3}$Department of Physics \& Astronomy, University of California, Riverside, CA 92521, USA\\
$^{4}$Institute for Astronomy, University of Edinburgh, Blackford Hill, Edinburgh, EH9 3HJ, UK\\
$^{5}$INAF-Osservatorio Astronomico di Trieste, via G. Tiepolo 11, Trieste, Italy\\
$^{6}$Scuola Normale Superiore, P.zzadei Cavalieri, I-56126 Pisa, Italy\\
$^{7}$IFPU–Institute for Fundamental Physics of the Universe,via Beirut 2, I-34151 Trieste, Italy\\
$^{8}$Department of Astronomy, University of Geneva, Chemin Pegasi 51, 1290 Versoix, Switzerland\\
$^{9}$Institute for Theoretical Physics, Heidelberg University, Philosophenweg 12, D-69120, Heidelberg, Germany\\
$^{10}$Max Planck Institut f\"{u}r Astronomie, K\"{o}nigstuhl 17, D-69117, Heidelberg, Germany\\
$^{11}$Tata Institute of Fundamental Research, Homi Bhabha Road, Mumbai 400005, India\\
$^{12}$Kavli Institute for Cosmology and Institute of Astronomy, Madingley Road, Cambridge, CB3 0HA, UK\\
$^{13}$Research School of Astronomy and Astrophysics, Australian National University, Canberra, ACT 2611, Australia\\
$^{14}$MIT Kavli Institute for Astrophysics and Space Research, 77 Massachusetts Ave., Cambridge, MA 02139, USA\\
$^{15}$Dipartimento di Fisica, Università di Trieste, Sezione di Astronomia, Via G.B. Tiepolo 11, I-34131 Trieste, Italy
}

\date{Accepted XXX. Received YYY; in original form ZZZ}

\pubyear{2023}

\begin{document}
\label{firstpage}
\pagerange{\pageref{firstpage}--\pageref{lastpage}}
\maketitle
\defcitealias{chen}{C17}
\defcitealias{cooper}{C19}
\defcitealias{becker2019}{B19}

\begin{abstract}
Intervening metal absorbers in quasar spectra at $z>6$ can be used as probes to study the chemical enrichment of the Universe during the Epoch of Reionization (EoR). This work presents the comoving line densities ($dn/dX$) of low ionization absorbers, namely, \ion{Mg}{II} (2796\AA), \ion{C}{II} (1334\AA) and \ion{O}{I} (1302\AA) across $2<z<6$ using the E-XQR-30 metal absorber catalog prepared from 42 XSHOOTER quasar spectra at $5.8<z<6.6$. Here, we analyse 280 \ion{Mg}{II} ($1.9<z<6.4$), 22 \ion{C}{II} ($5.2<z<6.4$) and 10 \ion{O}{I} ($5.3<z<6.4$) intervening absorbers, thereby building up on previous studies with improved sensitivity of 50\% completeness at an equivalent width of $W>0.03$\AA. For the first time, we present the comoving line densities of 131 weak ($W<0.3$\AA) intervening \ion{Mg}{II} absorbers at $1.9<z<6.4$ which exhibit constant evolution with redshift similar to medium ($0.3<W<1.0$\AA) absorbers. However, the cosmic mass density of \ion{Mg}{II} – dominated by strong \ion{Mg}{II} systems – traces the evolution of global star formation history from redshift 1.9 to 5.5. E-XQR-30 also increases the absorption pathlength by a factor of 50\% for \ion{C}{II} and \ion{O}{I} whose line densities show a rising trend towards $z>5$, in agreement with previous works. In the context of a decline in the metal enrichment of the Universe at $z>5$, the overall evolution in the incidence rates of absorption systems can be explained by a weak – possibly soft fluctuating – UV background. Our results, thereby, provide evidence for a late reionization continuing to occur in metal-enriched and therefore, biased regions in the Universe.
\end{abstract}




\section{Introduction}

The Epoch of Reionization (EoR) marks a major transition in the evolutionary history of the Universe when cosmic neutral hydrogen was (re)ionized by ultraviolet radiation from the first light sources and thereby marked an end to the Dark Ages. Studying reionization involves the understanding of the nature, formation and evolution of the first generation of stars and galaxies, quasars and the nature of the ionizing radiation. The theoretical and observational consensus view on the process of EoR indicates a patchy reionization with photoionized bubbles starting out in the vicinity of ionizing sources, which then expand and overlap, thus ionizing the whole Universe \citep{Loeb,oppenheimer2009,furlanetto,becker2015,bosman2022}. Studies using Lyman $\alpha$ optical depth measurements in high redshift quasar spectra \citep{fan2006, becker2015b, eilers2018, Choudhury2021, yang2020, bosman2022}, dark gap statistics in Lyman $\alpha$ forest \citep{songaila, furlanetto2004, gallerani, gnedin2017, nasir, zhu, zhu2022} and Lyman $\alpha$ damping wing absorption \citep{davies2018, banados2018, wang2020, greig2022} propose a late end to the EoR towards $z<6$. The same sources of ionizing radiation polluted the Universe with elements heavier than hydrogen and helium produced during stellar nucleosynthesis by ejecting them into the surrounding interstellar, circumgalactic and intergalactic media (ISM, CGM and IGM) through stellar and galactic feedback processes.

The CGM is the multi phase gas surrounding galaxies that extends from their disks to the virial radii and acts as a resource for star formation fuel and a venue for galactic feedback and recycling \citep{cgm}. Metals in the CGM offer many important insights into the evolutionary histories of their host galaxies. The quantity of metals depends on the past rate of metal ejection by outflows, and the ionization state of the metals can provide a key probe of the ionizing photon background. At large radii, the CGM is ionized by UV radiation from external sources rather than from the host galaxy: the metals in the diffuse medium will exhibit a transition in their ionization state with respect to the intensity of the ionizing background. Therefore, metal absorbers in galaxy halos at high redshifts can be used to study the ionization state of the gas near the EoR. Moreover, the ionizing ultra-violet (UV) background evolves to a softer spectrum with redshift as the population of luminous active galactic nuclei (AGN) decline towards redshift $z>5$ \citep{beckerbolton, D'Aloisio2017, kulkarni2019, faucher2020}. As a result, we would expect the outer parts of the CGM that are exposed to the ionizing background to undergo a transition in their ionization state. 

Quasar absorption spectroscopy is an effective method for detecting absorbers that reside in low-density gas such as CGM and IGM and at high redshift that are beyond the detection threshold of emission line surveys \citep{becker2015,peroux}.

The observations independently probe the global star formation history \citep{madau-dickinson, menard2011, matejek, chen}. Conventionally, the metal abundance of a galaxy or gas cloud is expressed in metallicity, given by the ratio of metals to neutral hydrogen. However, at $z>5$, due to the saturation of the Lyman $\alpha$ forest, it is difficult to measure \ion{H}{i} absorbers in the quasar spectra. As a result, observers use low ionization metal absorbers as probes for neutral hydrogen at high z. They are metal absorbers with ionization potentials less than neutral hydrogen (13.6 eV) (e.g., \ion{O}{i}, \ion{C}{ii}, \ion{Si}{ii}, etc.) and, therefore, can be used to trace neutral regions of CGM or IGM \citep{furlanetto, oppenheimer2009, keating2014, finlator2016}.

Studies of \ion{Mg}{ii} absorbers at redshifts $z<2$ based on their equivalent widths show that medium (0.3\AA\ $<W<$ 1.0\AA) and strong ($W>1.0$\AA) \ion{Mg}{ii} trace cool regions of outflows from blue star forming galaxies and thereby following the global star formation history \citep{zibetti2007, lundgren2009, weiner2009, noterdaeme2010, rubin2010, bordoloi2011, menard2011, nestor2011, prochter}. Meanwhile, works by \citet{kapzak2011, kapzak2012, nielsen2015} indicate that weak \ion{Mg}{ii} absorbers ($W<0.3$\AA) trace the corotating and infalling gas in the CGM. There can be significant variations in star formation rates, metal enrichment rate and halo assembly with lookback time and therefore it is necessary to extend the studies of \ion{Mg}{ii} systems beyond the peak of cosmic star formation rate at $z\sim2.5-3$ \citep{madau-dickinson} to understand galaxy transformation across different epochs. The first high redshift $z\sim6$ survey of \ion{Mg}{ii} was conducted by \citet{matejek} over $2.5<z<6$ using the Folded-port InfraRed Echellette (FIRE) spectrograph on Magellan. Further work by \citet{chen} added sightlines across $2<z<7$. Both works show that the incidence rates of strong \ion{Mg}{ii} absorbers decline with redshift following the cosmic star formation history while those of weaker absorbers remain constant with increasing redshift. 

Meanwhile, several works were conducted on $z<3$ weak \ion{Mg}{ii} absorbers after their detection by \citet{tripp} and \citet{churchill} to study their properties \citep[see][]{Rigby2002, Churchill2005,lynch, narayanan, 2008Narayanan, mathes2018, muzahid}. These low redshift studies on weak \ion{Mg}{ii} inferred that they arose from sub-Lyman Limit systems (sub-LLS having log $N_\ion{H}{I}<17.2$) with super-solar metallicity. \citet{mathes2018} predicted that the weak \ion{Mg}{ii} absorbers should not be detected at $z>3.3$ based on their number density statistics. Nevertheless, \citet{chen} detected some weak \ion{Mg}{ii} absorbers at $2<z<7$, but with low completeness, demonstrating the potential for near-IR instruments with improved sensitivity and resolution to detect more weak \ion{Mg}{ii} systems. 

 Theoretical studies on $z>5$ have indicated that dense neutral regions that have been polluted by metals will give rise to forests of low ionization absorption lines such as \ion{C}{ii} and \ion{O}{i} \citep{oh}. \citet{furlanetto2003} using their supernova wind model, argued that a substantial fraction of the metals in high redshift galactic superwinds which polluted the IGM existed as low ionization absorbers like \ion{C}{ii}, \ion{O}{i}, \ion{Si}{ii} and \ion{Fe}{ii}. Using a self-consistent multi-frequency UV model involving well constrained galactic outflows, \citet{finlator2015} shows that the \ion{C}{ii} mass fraction should drop towards lower redshifts relative to the increase in \ion{C}{iv} towards $z\la6$. Simulations to model the reionization of the metal-enriched CGM at $z\sim6$ in \citet{keating2014} and \citet{doughty} predict a rapid evolution of \ion{O}{i} at $z>5$ that probe regions of neutral hydrogen at these redshifts. 

Observations by \citet{cooper} using 69 intervening systems from Magellan/FIRE and Keck/HIRES find that the column density ratios of \ion{C}{ii}/\ion{C}{iv} increase towards $z>5$ which could be due to the combined effect of lower chemical abundance and softer ionizing background at $z\sim6$. Similar results can be seen in \citet{becker2006, becker2011} where there is a high number density of low-ionization absorbers such as \ion{C}{II} and \ion{O}{i} at $z\sim6$. The first self consistent survey for \ion{O}{i} absorbers using a larger number of sightlines (199) on Keck/ESI and VLT/XSHOOTER was conducted by \citet{becker2019}. They found an upturn in the \ion{O}{i} comoving line density at $z>5.7$, which they interpreted as an evolution in the ionization of metal-enriched gas, with lower ionization states being more common at $z\sim6$ than at $z\sim5$. Additionally, studies on \ion{C}{iv} at $z>5$ \citep[e.g.,][]{dodorico2010, dodorico2013, Davies2023} show the declining trend in absorption path density or mass density of highly ionized carbon with increasing redshift. \citet{finlator2016}, comparing the existing observational data from \citet{becker2011} and \citet{dodorico2013} with their models for different UV backgrounds, demonstrate that a softer fluctuating UV background reproduces the observed \ion{Si}{iv}/\ion{C}{iv} and \ion{C}{ii}/\ion{C}{iv} distributions.

In general, the increasing trend observed in low ionization absorbers and the decline in the incidence rates of high ionization absorbers point to a phase transition occurring in the CGM and IGM at $z>5$. Considering the metal absorbers as potential tracers of the host galaxies from which they are ejected, they can provide information about the relation between the galaxies and the absorbers, the environments in which they arise as well as the factors that drive their evolution. Furthermore, reionization models have not been fully successful in reproducing the observed trends in metal absorbers redshift evolution \citep[e.g.,][]{keating2016, finlator2016, doughty}. More observational constraints along with cosmological hydrodynamic simulations are required to improve our understanding of the reionization history of the Universe.

In particular, deeper observations over a larger number of sightlines are required to better characterize the evolution of metal absorbers and to understand how the metal content and ionization state of galaxy halos are impacted by reionization. This paper uses data from the ESO VLT Large Program - the Ultimate XSHOOTER legacy survey of quasars at $z\sim5.8-6.6$ \citep[XQR-30, ][]{xqr30}. XQR-30 is a spectroscopic survey of 30 (+12 from the archive and therefore, extended XQR-30 or E-XQR-30) quasars at high redshifts in the optical and near-infrared wavelength range (3000 - 25000 \AA) using the XSHOOTER spectrograph \citep{vernet} at the ESO Very Large Telescope (VLT). The archival observations of quasars from XSHOOTER have the same magnitude and redshift range as that of XQR-30 quasars with similar resolution and signal-to-noise (SN) ratio. The E-XQR-30 quasar spectra have an intermediate resolution of $R\sim10,000$ and a minimum (median) signal-to-noise (S / N) ratio of 10 ($\sim$29) per 10 \kms\ spectral pixel at a rest wavelength of 1258\AA\ \citep{xqr30}. The primary aim of this work is to understand the evolution of CGM absorbers at $z>5$ and the nature, particularly the strength, of ionizing photons towards the tail end of the EoR. This study is also important in the light of the recent revival of the possibility of quasars contributing significantly to the UV background at $z\sim6$ \citep{grazian2023, harikane2023, maiolino2023}. E-XQR-30 has already been used to constrain the end of EoR \citep{bosman2022,zhu}, studying the early quasars and their environment \citep{bischetti2022} as well as the evolution of high ionization absorbers like \ion{C}{iv} \citep{Davies2023}.

The metal absorber data used in this study are obtained from the E-XQR-30 metal absorber catalog \citep{xqr30catalog}. The catalog has substantially increased the sample size especially for \ion{C}{ii} and weak \ion{Mg}{ii} and the pathlength for the high redshift metal absorbers such as \ion{O}{i} and \ion{C}{ii} with deeper observations on a large number of objects and improved sensitivity of the XSHOOTER spectrograph compared to previous high-$z$ metal absorber surveys. This paper mainly focuses on low ionization absorbers such as \ion{Mg}{ii}, \ion{C}{ii} and \ion{O}{i}, including the first statistical study of weak Mg II absorbers at $2<z<6$, and studies their redshift evolution to characterise galaxy transformation across different epochs. The results from this work are also compared to previous survey results to understand how the increased sensitivity and resolution towards high redshift improves our understanding of the cosmic evolution of the absorbers. 

The outline of the paper is as follows: Section \ref{sec:Methods} describes the E-XQR-30 metal absorber catalog from which the data for this work are obtained and the techniques used to study the evolution of metal absorbers. We present the results of the cosmic evolution for each of the ions and their comparison with previous works in Section \ref{sec:Results}. The impacts of the observed trends in the absorber number densities with redshift are discussed in Section \ref{sec:Discussion}. Finally, a short summary of the entire paper is given in Section \ref{sec:Conclusions}. Throughout this work, we adopt the $\Lambda$ CDM cosmology with $H_0=67.7$\kms Mpc$^{-1}$ and $\Omega_\text{m}=0.31$ \citep{refId0}.



\section{Methods}
\label{sec:Methods}

\subsection{The E-XQR-30 Metal Absorber catalog}
\label{subsec:The E-XQR-30 Metal Absorber catalog}

 The metal absorber catalog prepared by \citep{xqr30catalog} used quasar spectra from E-XQR-30 consisting of 42 quasars \citep{xqr30}. All details related to the catalog can be found in \citet{xqr30catalog} and this section summarises some key features relevant to this work. After applying the data reduction procedures for the 42 quasar spectra as outlined in \citet{xqr30}, the spectra from each of the spectroscopic arms of XSHOOTER (VIS and NIR) were combined together into a single spectrum for each quasar. The adopted emission redshifts of each quasar were calculated from emission lines where available or from the apparent start of Lyman alpha forest as listed in Table B1 in \citet{xqr30catalog}. For the preparation of the metal absorber catalog, only absorbers redward of the Lyman $\alpha$ emission line and to a maximum redshift of 5000\kms\ below the quasar emission redshift, were considered. The quasar spectra is completely absorbed by the Lyman $\alpha$ forest due to intervening neutral hydrogen so it is very challenging to search for absorption lines at wavelengths shorter than Lyman $\alpha$. The wavelength regions affected by skyline or telluric contamination were excluded from the absorber search. Absorbers found in spectral regions affected by broad absorption line (BAL) troughs \citep{bischetti2022, bischetti2023} were flagged in the catalog. 

The metal absorption catalog has been prepared using an automated search for candidate systems, checking for spurious detections through customised algorithms and visual inspection and fitting of the lines with Voigt profile to obtain column density (log $N$) and Doppler ($b$) parameter. Many of the procedures used \textsc{ASTROCOOK} \citep{cupani}, a Python software for detecting and fitting quasar absorption lines. 

The metal absorber components were grouped into systems using a similar method outlined in \citet{d'odorico2022} where the components that are separated by less than 200 \kms\ were combined into a single system using an iterative method through the list of the systems for each line of sight. For each system, the total rest frame equivalent width $W$ and the $W$-weighted mean redshift from the constituent components were measured from the best-fit Voigt profiles.

The absorbers are classified as proximate or intervening based on their velocity separation from quasar redshifts to avoid absorbers with ionization states or abundance patterns different from the intrinsic absorber population due to close proximity to the quasar. Based on the work of \citet{perrotta}, the E-XQR-30 metal absorber catalog adopted a minimum velocity separation of 10,000 \kms\ from the quasar redshift to label an absorber as intervening (also known as non-proximate) along the line of sight. In this case, the maximum redshift at which an intervening ion can be detected is 
\begin{equation}
    z_{\text{max}}=(1+z_{\text{em}})\times \text{exp}(-10,000\ \text{\kms}/c)-1,
\end{equation} where $z_{\text{em}}$ is the emission redshift of the quasar and $c$ is the speed of light.
Although the primary sample\footnote{The primary sample consists of only automatically detected systems whose completeness and false positives are well constrained.} of the E-XQR-30 catalog adopts 10,000 \kms\ as the proximity limit, the published data allow users to set a different velocity threshold if desired. To this end, the work presented here also adopts the limits of 3000 \kms and 5000 \kms to compare directly with previous works on \ion{Mg}{ii} and \ion{O}{I} in the literature. 

The evolution of the absorbers in the CGM across cosmic time can be studied by using either the individual absorption component or system data for each ion. For the purpose of this paper, only system data are considered for the redshift evolution studies of the absorbers, so that our results can be compared with previous works such as \citet{chen,cooper} and \citet{becker2019} which have different spectral resolutions other than that of XSHOOTER \citep{d'odorico2022, xqr30catalog}. 

The metal absorber catalog consists of 778 systems in total including 280 \ion{Mg}{ii}, 22 \ion{C}{ii} and 10 \ion{O}{i} intervening systems, providing a significant increase in the number of high redshift absorbers with high spectral resolution and S/N ratio compared to previous surveys. For example, the catalog almost doubled the number of \ion{C}{II} absorbers at $z>5$ and detected a substantial population of 138 weak \ion{Mg}{ii} absorbers at $z>2$. The E-XQR-30 sample also increases the absorption pathlength for \ion{C}{ii} and \ion{O}{i} absorbers by 50\% \citep{xqr30catalog} at $5.17<z<6.38$ in comparison to other works in the literature.

\subsection{The low-ionization absorber line statistics: dn/dX }
\label{subsec: The low-ionization absorbers: dn/dX}

We focus on studying the evolution of low ionization absorbers, namely, \ion{Mg}{ii}, \ion{C}{ii} and \ion{O}{i}, across redshift using E-XQR-30 metal absorber catalog. In order to study the change in the metal content of the galaxy halos across redshift, we use the absorption path density ($dn/dX$), also known as comoving line density. \footnote{The term 'comoving line density' in this work is same as 'number density', 'line density' and 'incidence rates' used in other publications.} It gives the number of absorbers per unit absorption path length interval. The absorption path for a given redshift is 
\begin{equation}\label{eq:1}
    X(z)=\frac{2}{3\Omega_m}(\Omega_m(1+z)^3+\Omega_\Lambda)^{1/2}
\end{equation} \citep{bahcall} where $\Omega_m$ is matter density and $\Omega_\Lambda$ is dark energy density parameter of the Universe. The quantity $dn/dX$ also normalises out the redshift dependence of the occurrence of the non-proximate absorbers along the line of sight. If $dn/dX$ is flat, it indicates no comoving evolution meaning that the the product of absorber cross section and comoving volume density is fixed for a population of absorbers. 

We calculate the $dn/dX$ following the steps outlined in Section 6.3 of \citep{xqr30catalog}. For each ion, the intervening absorbers in the primary sample are binned into different redshift intervals in such a way that each redshift range covers similar pathlengths ($\Delta X=X(z_2)-X(z_1)$). The number of redshift bins for each ion is determined by ensuring that there is sufficient number of absorbers in each bin. The proximity limit used in this work is 10,000 \kms unless otherwise specified. 

The survey completeness plays a major role in statistical analysis of the absorbers. \citet{xqr30catalog} have characterised the completeness for the E-XQR-30 sample by creating 20 mock spectra for each quasar in the survey for which the absorber properties are known beforehand. These spectra are then processed in a similar way as the actual spectra to estimate the completeness as a function of the equivalent width, redshift, column density and $b$ parameters for each of the ions. The sample completeness reaches 90\% at $W=0.09$\AA\ and 50\% at $W=0.03$\AA. For this study, we only retain absorbers with $W>0.03$\AA\ and apply the completeness correction as a function of the equivalent width as follows:
\begin{equation} \label{eq:2}
    \text{Completeness}(W) = S_y(\text{arctan}(S_xW+T_x)+T_y)
\end{equation} where $S_x=59.5$, $S_y=0.39$, $T_x=-1.55$ and $T_y=1.01$ \citep[also see Figure 8 in][]{xqr30catalog}. 

To calculate the completeness-corrected $dn/dX$, the absorbers in each redshift interval are split into $W$ bins of width 0.03\AA. For each $W$ bin, the number of absorbers is divided by the completeness correction (equation \ref{eq:2}) and these values are summed to get the total completeness-corrected number of absorbers in each redshift range. This method is applied to all absorbers studied in this work to correct for completeness unless otherwise specified, and rest frame W values are used throughout. The completeness correction can also be applied to the absorbers as a function of column density \citep[see Figure 8 and Table 3 in][]{xqr30catalog}; however, both methods give consistent results. 

After obtaining the completeness corrected counts in each redshift bin, the absorption path length interval, $\Delta X$ corresponding to each of those bins are calculated using the Python code published with \citet{xqr30catalog}\footnote{\url{https://github.com/XQR-30/Metal-catalogue/tree/main/AbsorptionPathTool}}, which removes masked regions from the absorption path. 

Once the $dn/dX$ values are calculated, the errors associated with $dn/dX$ are computed using a Poisson distribution approximation for the absorber counts. The $1\sigma$ confidence limits of the $dn/dX$ values are calculated using equations (9) and (14) from \citet{gehrels} for the upper and lower limits, respectively. The upper limit is calculated using
\begin{equation}
\label{eq:upper error}
    \lambda_u=(n+1)\Bigg[1-\frac{1}{9(n+1)}+\frac{S}{3\sqrt{n+1}}\Bigg]^3
\end{equation}.
The $1\sigma$ lower limit is estimated using
\begin{equation}
\label{eq:lower error}
    \lambda_l=n\bigg(1-\frac{1}{9n}-\frac{S}{3\sqrt{n}}+\beta n^\gamma\bigg)^3
\end{equation} where n is the number of absorbers. The parameter values of S, $\beta$ and $\gamma$ are taken from Table 3 of \cite{gehrels} corresponding to 1 sigma (0.8413) confidence limits. While calculating the errors, the completeness correction factor was also considered since $dn/dX$ involves completeness corrected counts for each redshift interval. The $dn/dX$ values are then plotted against path length weighted mean redshift (\textlangle z\textrangle) to see the cosmic evolution of the absorber.
\section{RESULTS}
\label{sec:Results}
The following sections show how the comoving line density $dn/dX$ of each ion evolves with cosmic time and the effects of improved spectral resolution on the results compared with earlier works.

\subsection{Mg {\scriptsize II} \texorpdfstring{$\lambda\ 2796$}{lone}\AA\ \& \texorpdfstring{$\lambda\ 2803$}{ltwo}\AA}
\label{subsec: MgII}

\ion{Mg}{ii} traces both neutral and ionized gas in metal-enriched galaxy halos. Analysing the evolution of \ion{Mg}{ii} absorbers at high redshifts furthers our understanding of the mechanisms through which the halos were populated with \ion{Mg}{ii} up to the peak of cosmic star formation at $z\approx2$ \citep{matejek}. The E-XQR-30 metal absorber primary catalog has 280 intervening \ion{Mg}{ii} systems among which 264 were detected with $W>0.03$\AA\ in the redshift range 1.944 -- 6.381 along a path length of $\Delta X=553.2$. Only systems with $W>0.03$\AA\ are used to analyse their redshift evolution.

\begin{figure}
    \includegraphics[width=\columnwidth]{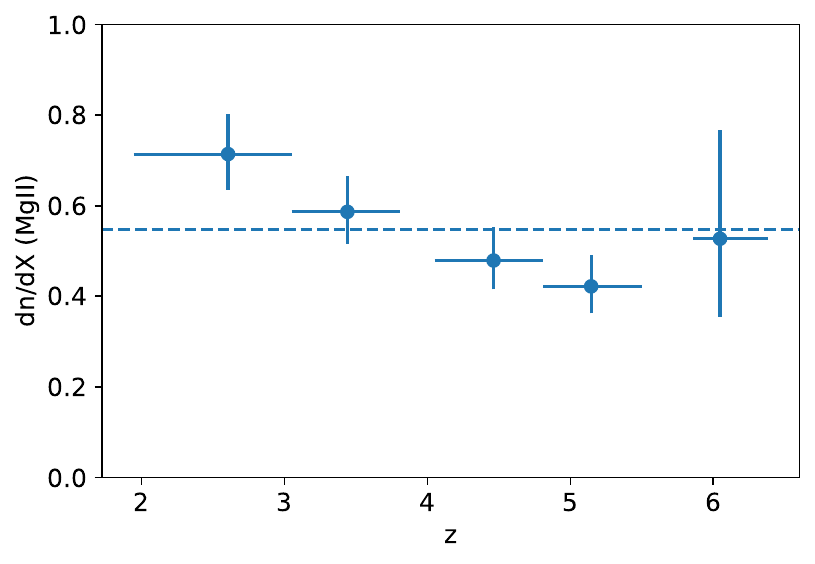}
    \caption{Evolution of the comoving line density of \ion{Mg}{ii} absorbers with redshift using a proximity limit of 10,000 \kms. The comoving line density of \ion{Mg}{ii} declines with redshift until $z\sim5$ after which it is associated with larger errors. The normalized mean $dn/dX$ value is denoted by a dashed horizontal line. For comparison with literature measurements at $z<2$, see Figure \ref{fig:2}.}
    \label{fig:1}
\end{figure}

The \ion{Mg}{ii} absorbers are binned into five redshift intervals (see Table \ref{tab:1}) in such a way that they cover similar absorption pathlengths with the exception of the highest redshift bin ($\Delta X=17.76$). The redshift regions $3.81 < z < 4.05$ and $5.5 < z < 5.86$ are excluded because they correspond to contaminated wavelength regions where far fewer absorbers can be robustly measured. The $dn/dX$ values obtained are shown in Table \ref{tab:1} and the trend in $dn/dX$ across redshift $z$ is shown in Figure \ref{fig:1}.
\begin{table*}
    \centering
    \caption{The $dn/dX$ and $\Omega$ values of \ion{Mg}{ii} absorber systems using a proximity limit of 10,000 \kms\ after binning them in five redshift intervals. It can be seen that certain redshift intervals are masked. Counts are corrected for completeness using equation \ref{eq:2}.}
    \label{tab:1}
    \begin{tabular}{|l|c|c|c|c|c|c|c|c|}
        \hline
		$z$ range & \textlangle $z$\textrangle & $\Delta X$ & counts & \multicolumn{1}{|p{1cm}|}{\centering corrected\\counts} & $dn/dX$ & \multicolumn{1}{|p{1.5cm}|}{\centering $\delta(dn/dX)$\\(+,-)} & $\Omega \times10^{-8}$ & $\delta\Omega\times10^{-8}$\\
            \hline
		1.944-3.050 & 2.60 & 119.18 & 81 & 85.09 & 0.71 & 0.09, 0.08 & 14.39 & 8.33\\
		3.050-3.810 & 3.44 & 118.73 & 68 & 69.63 & 0.59 & 0.08, 0.07 & 4.01 & 1.41\\
		4.050-4.810 & 4.46 & 123.71 & 56 & 59.26 & 0.48 & 0.07, 0.06 & 2.39 & 1.13\\
		4.810-5.500 & 5.15 & 123.18 & 50 & 51.97 & 0.42 & 0.07, 0.06 & 3.55 & 1.57\\
		5.860-6.381 & 6.05 & 17.76 & 9 & 9.36 & 0.53 & 0.24, 0.17 & 0.26 & 0.12\\
		\hline
    \end{tabular}
    
\end{table*}

It is evident from Figure \ref{fig:1} that \ion{Mg}{ii} absorbers decline in comoving line density with increasing redshift. A slight increase is seen at the highest redshift but it is associated with a large error due to few counts and small absorption path in the final bin. Therefore, the increase from $\langle z\rangle \sim5$ to $\langle z\rangle \sim6$ is not statistically significant given the errors.

It is also interesting to see how the \ion{Mg}{ii} absorbers evolve with redshift if they are sub-divided based on their strength. Previous studies of strong \ion{Mg}{ii} absorbers indicate that they trace global star formation history through galactic outflows and weak \ion{Mg}{ii} systems trace the accreting and co-rotating gas in galaxy halos (see Sections \ref{subsec:weak MgII absorbers at high z} and \ref{subsec:strong MgII}). \citet{chen} (\citetalias{chen} from here on) is a large survey of high redshift \ion{Mg}{ii} absorbers using 100 quasars at $3.55\le z\le7.09$ with the Magellan/FIRE spectrometer detecting 280 \ion{Mg}{ii} absorbers. \citetalias{chen} analysed the evolution of medium and strong \ion{Mg}{ii} absorbers by applying a proximity limit of 3000\kms.  

\begin{figure}
    \centering
    \includegraphics[width=\columnwidth]{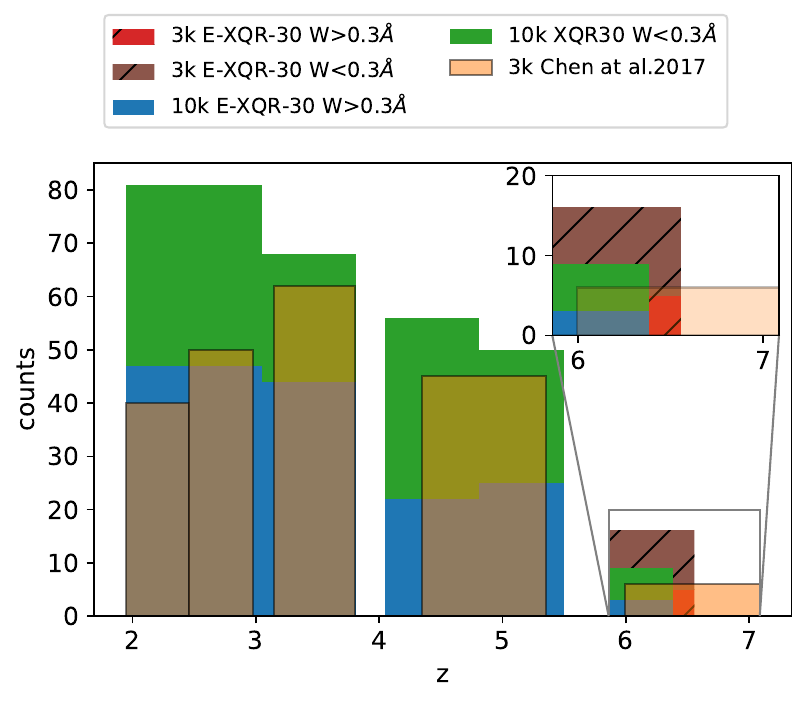}
    \caption{The number of intervening \ion{Mg}{ii} absorber systems in different redshift intervals from E-XQR-30 and \citetalias{chen}. The E-XQR-30 sample is stacked on the basis of their rest frame equivalent widths and only absorbers above 50\% completeness limit are shown here. The green bins show the weak absorbers, and the blue bins give the medium and strong absorber counts with a proximity limit of 10,000 \kms. As shown in the inset, the brown hatched bins give the new number of weak absorbers  and the red hatched bins give the sum of medium and strong absorber counts when a smaller value of proximity limit (3000 \kms) is applied. It can be seen that when the proximity limit is reduced, it increases the width of the highest redshift bin and the number of $W<0.3$\AA\ and $W>0.3$\AA\ absorbers. The absorber counts from \citetalias{chen}, which uses a proximity limit of 3000 \kms, are shown in a light orange color. The medium and strong absorbers ($W>0.3$\AA) from both works show a decline in the number counts with redshift while the weak absorbers ($W<0.3$\AA) almost remain constant with redshift except for the highest redshift bin. The masked redshift regions for both E-XQR-30 and \citetalias{chen} are denoted by the gaps in the histogram.}
    \label{fig:3}
\end{figure}

E-XQR-30 has detected 66 out of 70 \ion{Mg}{ii} absorber systems from the 19 quasars that were used in \citetalias{chen} because one of the missing systems falls in a noisy region and the other three systems were observed to be better explained by other ion transitions at various redshifts. In addition, the survey enabled the detection of 95 additional systems in those quasars due to the improved senistivity. \citep{xqr30catalog}. A histogram of the number counts of the intervening \ion{Mg}{ii} absorbers with different strengths used for the $dn/dX$ analysis between this work and \citetalias{chen} across different redshift intervals is shown in Figure \ref{fig:3}. The E-XQR-30 sample is divided into two groups; one group consisting of only weak absorbers ($W<0.3$\AA) and the other group with both medium and strong absorbers ($W>0.3$\AA) for better comparison with \citetalias{chen}. Although \citetalias{chen} were able to detect some weak systems, the overall completeness of their data was not enough to produce robust statistical data. We present for the first time a large population of weak intervening \ion{Mg}{ii} (123 absorbers after applying the 50\% completeness cut and masking the contaminated redshift intervals) at 2<z<6 sufficient for a statistical analysis. For E-XQR-30, the weak absorbers are stacked on top of the absorbers with $W>0.3$\AA\, shown in green and blue bins respectively. The number of absorbers from \citetalias{chen} is shown in light orange. It is evident from the histogram that E-XQR-30  has a larger total sample available for analysis after applying the completeness limit and a larger proximity limit of 10,000 \kms. However, \citetalias{chen} has a slightly bigger sample if only absorbers with $W>0.3$\AA\ (medium and strong) absorbers are considered due to their larger number of background quasars. The histogram shows a wider bin at the highest redshift interval for \citetalias{chen} compared to the E-XQR-30 bin because the former observed quasars that covered broader redshift ranges. The effect of using a smaller proximity limit on the sample size and the total redshift range covered is also investigated in this work. When the proximity limit for E-XQR-30 is changed to 3000 \kms, the upper limit of the redshift bin changes from 6.381 to 6.555 and the number of absorbers in the last redshift bin increases slightly, as shown by the hatched brown and red coloured bins in the inset. There is a total increase of 7 absorbers: 5 weak and 2 medium. It should also be noted that the number of strong absorbers remained zero even after reducing the proximity limit. 

\begin{figure}
    \centering
    \includegraphics[width=\columnwidth]{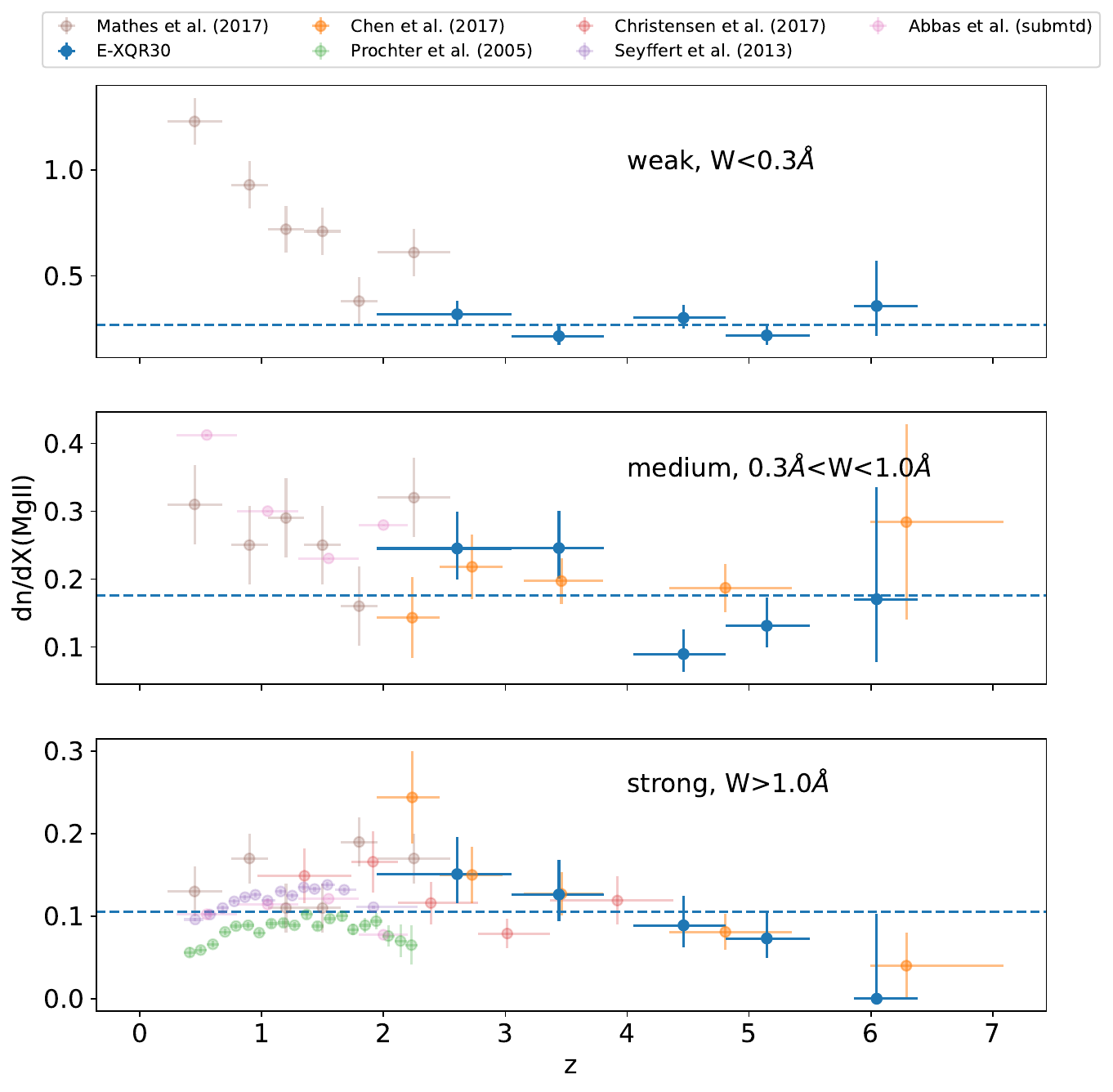}
    \caption{Redshift evolution of \ion{Mg}{ii} absorbers divided into three samples according to their equivalent widths and using 10,000 \kms as the proximity limit. The results from E-XQR-30 are shown in blue and those from \citetalias{chen} are given in orange. The $dn/dX$ values at lower redshift from various literature are also included in the figure. The top panel shows the evolution of weak \ion{Mg}{ii} absorbers ($W<0.3$\AA). The middle and bottom panels show the evolution of medium and strong \ion{Mg}{ii} absorbers, on which the $dn/dX$ values from \citetalias{chen} are also plotted. The blue dashed horizontal lines in each panel represent the pathlength weighed mean $dn/dX$ for E-XQR-30 sample. The $dn/dX$ of absorbers with $W<1.0$\AA\ show a flat evolution with redshift for $z>2$ while those with $W\ge1.0$\AA\ show a decline in comoving line density towards $z>2$. }
    \label{fig:2}
\end{figure}

 Figure \ref{fig:2} shows the evolution of the \ion{Mg}{ii} comoving line density with redshift for different strengths of the absorbers. The absorbers are subdivided based on their equivalent widths: $W<0.3$\AA, $0.3<W<1.0$\AA\ and $W>1.0$\AA. The $dn/dX$ values for \ion{Mg}{ii} absorbers with different equivalent widths and redshifts can be found in Table \ref{tab:4}. There are three interesting findings worthy of further discussion: the remarkably high $dn/dX$, the flat evolution of weak and medium ($W<1.0$\AA) systems with redshift at $2<z<5$, and the potential upturn at $z>5$.

As demonstrated in \citet{codoreanu} and \citetalias{chen}, associating each \ion{Mg}{ii} absorber with a single galaxy (at a limiting magnitude ($M_{AB}\le -17.5$) or halo mass cut (log $M_h>10.2$) respectively) cannot reproduce the high redshift $dn/dX$ to within a factor of 10 or more. Both studies adopt observationally derived scaling relations in which the CGM absorption radius scales with halo mass and the covering fraction for weak \ion{Mg}{ii} absorbers is greater than 80 percent within 50 kpc \citep{churchill2013,nielsen}. The superior spectral resolution and higher signal to noise ratio of our study that includes weak \ion{Mg}{ii} systems further underscores the tension since $dn/dX$ is dominated by weak systems (see y-axis of Figure \ref{fig:2}). \citetalias{chen} were able to reconcile the tension, at least with medium absorbers, by allowing the mass cut to vary with redshift, and further integrating down the galaxy mass function as $z$ approaches 6 to include galaxies with log $M_h>8$ ($=\text{log}\ M_*>7$).

The comoving line density of strong absorbers ($W > 1.0 $\AA) show a declining trend with redshift. Thus, the decrease in $dn/dX$ for the total sample (see Figure \ref{fig:1}) can be attributed to this decline to $z\sim5$, while the upturn (with high error bars) at the highest redshift bin could be due to the increase at $\langle z \rangle \sim 6$ in comoving line density of absorbers with $W < 0.3$\AA. This upturn is not statistically significant due to the short pathlength of this redshift bin. More observations focusing on redshifts $z>5$ must be made to reach robust conclusions.

The $dn/dX$ evolution of \ion{Mg}{ii} using the E-XQR-30 metal absorber catalog agrees with the results of \citetalias{chen} for medium and strong absorbers ($W>0.3$\AA) although there are differences in the pathlength weighted mean redshifts at which the $dn/dX$ values are calculated. The difference in the proximity limits used in both works must also be taken into consideration when comparing the results. Adopting a 3000 \kms\ proximity limit to match \citetalias{chen} analysis results in $dn/dX$ values similar to the 10,000 \kms\ (see Table \ref{tab:4}). Therefore, we chose to plot the primary E-XQR-30 catalog values in Figure \ref{fig:3}. The consistency in the comoving line density statistics regardless of the proximity limits used, shows that our measurements are robust to the choice of proximity zone limit.

This work is also consistent with the results from other \ion{Mg}{ii} comoving line density analyses such as \citet{matejek, codoreanu} and \citet{zou}. The work of \citet{matejek} is a precursor of \citetalias{chen} using 46 quasar spectra from FIRE. They observed no evolution for absorbers with 0.3\AA$<W<$1.0\AA\ while the $dn/dX$ of strong absorbers in their work showed a slight increase until $z\sim3$, which could be due to the small number of detections in the corresponding redshift intervals, after which they decline with redshift. \citet{codoreanu} detected 52 \ion{Mg}{ii} absorbers in the redshift range $2<z<6$ from high quality spectra of four high-$z$ quasars from XSHOOTER where they demonstrated a flat redshift evolution of $dn/dX$ for weak and medium absorbers although the evolution of strong absorbers was subjected to limited sample size. Also, \citet{zou} obtained the $dn/dX$ values of strong \ion{Mg}{ii} absorbers at $2.2<z<6.0$ using Gemini GNIRS which are consistent with the E-XQR-30 results. Furthermore, our remarkably flat $dn/dX$ for weak \ion{Mg}{II} absorbers provides context for the anticipated cross-correlation analysis of the \ion{Mg}{II} forest from JWST data \citep{hennawi}.

In Figure \ref{fig:2}, $dn/dX$ values from \ion{Mg}{ii} absorber surveys at lower redshift are also included to compare the absorber evolution at $z<2$ with the results from our work. \citet{prochter} studied strong \ion{Mg}{II} absorbers across $0.35<z<2.3$ and found that they roughly follow the global star formation rate density. Similar findings have been reported by \citet{seyffert2013} on strong \ion{Mg}{ii} which are found to increase by 45\% approximately from $z=0.4$ to $z=1.5$ associating these systems with outflows from star-forming galaxies. The $dn/dX$ from \citet{christensen2017} for strong systems at $0.9<z<4.4$ roughly agrees with the trends in previous high-resolution surveys including our work peaking at $\langle z\rangle\sim1.9$ and then declining towards high redshift. \citet{mathes2018} studied the evolution of \ion{Mg}{II} with equivalent widths $W>0.01$\AA\ at $0.1<z<2.6$ and thereby calculated the $dn/dX$ of weak, medium and strong absorbers. Recently, \citet{abbas} analysed the evolution of \ion{Mg}{ii} absorbers with $W>0.3$\AA\ using a new method to measure the column densities of \ion{Mg}{ii} systems using the Australian Dark Energy Survey (OzDES) over the redshift range of $0.33\le z\le 2.19$. The $dn/dX$ values from the above mentioned earlier works are colour coded accordingly as given in the legend at the top of the figure. 
At $z<2$, the weak systems are observed to increase towards the present epoch which can be attributed to the metallicity build-up and the decreasing intensity of the ionizing radiation in the CGM giving rise to more weak absorbers. The constant evolution of medium \ion{Mg}{ii} absorbers at z>2 extends to lower redshifts in such a way that the number density of the absorbers balances out the absorber cross section across the whole redshift range. However, strong \ion{Mg}{ii} absorber evolution, in general, traces the trend in global star formation history \citep{madau-dickinson} which rises across cosmic time until $z\sim2$ after which it shows a gradual decline towards the present epoch.

\begin{table*}
    \centering
    \caption{The $dn/dX$ values for \ion{Mg}{ii} absorbers based on the strength of the absorption profiles. A proximity limit of 10,000 \kms is used here and the grey shaded regions show the $dn/dX$ of \ion{Mg}{ii} absorbers for a proximity limit of 3000 \kms for comparison with \citetalias{chen}. There is no significant change when the proximity limit is changed. The completeness of the sample is equal to unity at $W>1.0$\AA.}
    \label{tab:4}
    \begin{tabular}{|l|c|c|c|c|c|l|}
    \hline
    $z$ range & \textlangle $z$\textrangle & $\Delta X$ & counts & \multicolumn{1}{|p{1cm}|}{\centering corrected\\counts} & $dn/dX$ & \multicolumn{1}{|p{1.5cm}|}{\centering $\delta(dn/dX)$\\(+,-)} \\
    \hline
    \multicolumn{7}{|c|}{$W_{\ion{Mg}{ii}}<0.3$\AA} \\
    \hline
    1.944-3.050 & 2.60 &  119.18  &  34 & 37.88 & 0.32 &         0.06      ,    0.05\\
    3.050-3.810 &  3.44 &  118.72  &  24 & 25.46 & 0.21 &        0.05    ,       0.04\\
    4.050-4.810 &  4.46 &  123.71  &  34 & 37.20 & 0.30 &        0.06        ,      0.05\\
    4.810-5.500  &  5.15 &  123.18 &   25 & 26.83 & 0.22 &         0.05            ,       0.04\\
    5.860-6.381 & 6.05 &   17.76 &   6 & 6.34 & 0.36 &       0.21                ,     0.14\\
    \rowcolor{lightgray}
    5.860-6.555 & 6.06  &   35.87 &   11 & 11.73 &0.33   &       0.13   ,        0.09\\

    \hline
    \multicolumn{7}{|c|}{$0.3$\AA$<W_{\ion{Mg}{ii}}<1.0$\AA} \\
    \hline
    1.944-3.050 & 2.60 &  119.18 &   29 & 29.21 & 0.25    &        0.05    ,      0.05\\
    3.050-3.810 &  3.44 &  118.73  &  29 & 29.17 & 0.25   &          0.05    ,      0.05\\
    4.050-4.810 &  4.46 &  123.71  &  11 & 11.05 &  0.09     &      0.04     ,     0.03\\
    4.810-5.500  &  5.15  & 123.18  &  16 & 16.14 & 0.13     &       0.04     ,     0.03\\
    5.86-06.381 & 6.05  &  17.76  &  3 & 3.02 &  0.17     &       0.17    ,       0.09\\
    \rowcolor{lightgray}
    5.860-6.555 & 6.06  &   35.87  &   5 & 5.04 & 0.14    &        0.09   ,       0.06\\
    \hline
    \multicolumn{7}{|c|}{$W_{\ion{Mg}{ii}}>1.0$\AA} \\
    \hline
    1.944-3.050 & 2.60 &  119.18  &  18 & 18 & 0.15     &       0.045    ,      0.04\\
    3.050-3.810  & 3.44 &  118.73  &  15 & 15 &  0.13    &        0.04    ,      0.03\\
    4.050-4.810 &  4.46 &  123.71  &  11 & 11 &  0.09     &      0.04    ,      0.03\\
    4.810-5.500  &  5.15 &  123.18   &  9 & 9 &  0.07    &       0.03     ,     0.02\\
    5.860-6.381 & 6.05  &  17.76   & 0 & 0 & 0            &       0.10      ,     0\\
    \rowcolor{lightgray}
    5.860-6.555 & 6.06  &   35.87 &    0 & 0   & 0      &          0.05    ,      0\\
    \hline

    \end{tabular}
\end{table*}

\begin{figure}
    \centering
    \includegraphics[width=\columnwidth]{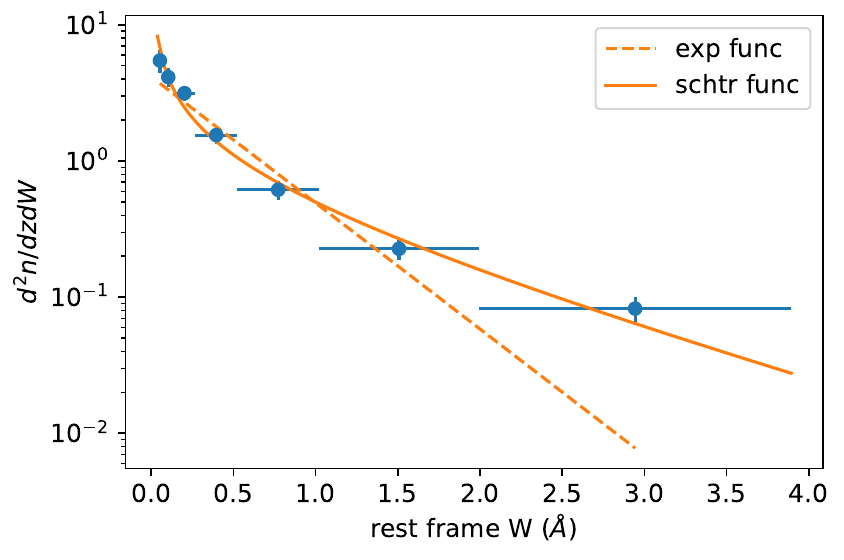}
    \caption{\ion{Mg}{ii} equivalent width distribution for the total sample corrected for completeness. The exponential function fit is shown in dashed orange and the Schechter function is shown in solid orange, which is a better fit for the equivalent width distribution.}
    \label{fig:total MgII W dist}
\end{figure}

Using a single sightline from deep XSHOOTER survey, \citet{bosman2017} detected 5 intervening \ion{Mg}{ii} systems at $z>5.5$ with 3 of them being weak absorbers showing that there is a possibility of steepening of equivalent width distribution at low equivalent widths. This is in agreement with the large number of detections of \ion{Mg}{ii} with $W<0.3$\AA\ at $z>2$ in this work. Using the weak absorbers sample, the prediction by \citet{bosman2017} can be verified by plotting the equivalent width distribution given by \begin{equation}
    \frac{d^2n}{dzdW} = \frac{N}{\Delta W \Delta z}
\end{equation} where $\Delta W$ is the equivalent width range and $\Delta z$ is the redshift pathlength. The blue points in Figure \ref{fig:total MgII W dist} indicate the $W$ distribution for the total sample of \ion{Mg}{ii} that are completeness corrected. 

\citet{nestor} has shown that equivalent width distribution for \ion{Mg}{ii} absorbers with $W>0.3$\AA\ can be fitted by an exponential function given by \begin{equation}
\label{eq:exp func}
    \frac{d^2n}{dzdW}=\frac{N^*}{W^*}e^{-W/W^*}
\end{equation} where $N^*$ is the normalisation factor and $W^*$ determines the exponential curve growth or decay. But \citet{bosman2017} showed that a single exponential function might not be the best fit over the whole range of equivalent width. We fit the $W$ distribution for the \ion{Mg}{ii} absorbers from this work at $1.9<z<6.4$ using equation \ref{eq:exp func} and the fit is indicated by the dashed orange curve in Figure \ref{fig:total MgII W dist}. The best fitting parameter values obtained are $W^*=0.47\pm0.04$ and $N^*=1.95\pm0.36$.  The function is a good fit for low equivalent widths but not for $W>1.0$\AA\ proving the prediction of \citet{bosman2017}. As a result, we then fit the distribution using a Schechter function of the form \begin{equation}\label{eq:schechter}
    \frac{d^2n}{dzdW} = \bigg(\frac{\Phi^*}{W^*}\bigg)\bigg(\frac{W}{W^*}\bigg)^\alpha e^{-W/W^*},
\end{equation} following the approach of \citet{kapzak&churchill, mathes2018} at $z<2$ and \citet{bosman2017} at $z>6$. Here, $\Phi^*$ is the normalisation factor, $\alpha$ is the low equivalent width power slope and $W^*$ is the turn over point where the low equivalent width power law slope shifts to an exponential cut-off. Using chi square statistic, the best fitting parameters are estimated to be $W^*=1.44^{+0.32}_{-0.26}$\AA, $\Phi^*=1.13^{+0.16}_{-0.15}$ and $\alpha=-0.66^{+0.09}_{-0.07}$. However, these values are different from those computed by \citet{bosman2017} for their sample of 3 absorbers at $5.9<z<7$. The fit using the Schechter function is represented in solid orange in Figure \ref{fig:total MgII W dist}, and provides a better fit than the exponential function. The redshift evolution of the equivalent width distribution is also explored in this work and the slope of the distribution is observed to steepen with redshift. More details can be found in Appendix \ref{MgII distribution with z}.

\subsection{C {\scriptsize II} \texorpdfstring{$\lambda\ 1334$}{lone}\AA}
\label{subsec: CII}

\ion{C}{ii} is one of the abundant ions present in the galaxy halos tracing the metal-enriched gas, where hydrogen is largely neutral. Since the first ionization energy of carbon (11.26 eV) is less than 13.6 eV, it will appear as singly ionized carbon (\ion{C}{ii}) in the otherwise neutral medium \citep{becker2015}. 

The primary E-XQR-30 absorber catalog contains 22 intervening \ion{C}{ii} absorbers observed along a path length of $\Delta X=77.8$ in the redshift interval $5.169<z<6.381$. Since there is only one strong \ion{C}{ii} transition in the searched spectral window, it is important to look for other low ionization transitions associated with \ion{C}{ii} such as \ion{O}{i} $\lambda1302$\AA, SiII $\lambda1260, 1526$\AA\ or Al II $\lambda1670$\AA\ to confirm the identification. The \ion{C}{ii} candidates are rejected if the probed spectral region reveals a non-detection of \ion{Si}{ii} and \ion{O}{i} because these associated ions are expected to have comparable equivalent widths \citep{becker2019, xqr30catalog}. 

\begin{figure}
    \centering
    \includegraphics[width=\columnwidth]{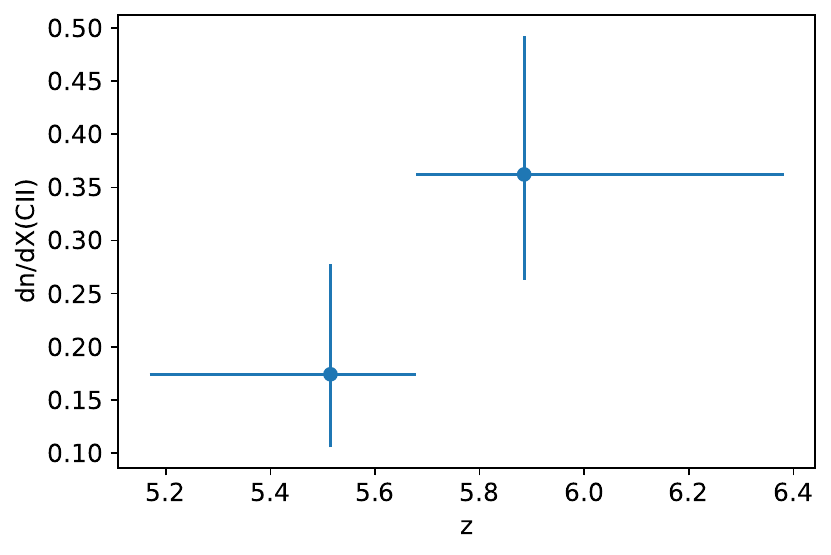}
    \caption{The evolution of \ion{C}{ii} absorbers across redshift applying a proximity limit of 10,000 \kms. The absorbers have $W>0.03$\AA. An increasing trend of comoving line density can be seen in the figure.}
    \label{fig:4}
\end{figure}

The comoving line density evolution of \ion{C}{II} is calculated using the same method applied to \ion{Mg}{II}. Only 19 \ion{C}{ii} systems with $W>0.03$\AA\ are considered. Due to the short path length available for \ion{C}{ii} detection redward of the saturated Lyman-alpha forest, we split the available redshift range into two bins with equal pathlengths, $\Delta X$. The redshift ranges, $dn/dX$ and the associated errors are given in Table \ref{tab:2}. The evolution of \ion{C}{ii} absorbers with redshift is shown in Figure \ref{fig:4}. The $dn/dX$ value for \ion{C}{ii} doubles from $\langle z\rangle$=5.5 to 5.9. 
\begin{table*}
    \centering
    \caption{The $dn/dX$ values of \ion{C}{ii} with the associated errors at different redshift intervals adopting a proximity limit of 10,000 \kms.}
    \label{tab:2}
    \begin{tabular}{|l|c|c|c|c|c|r|}
        \hline
		$z$ range & \textlangle $z$\textrangle & $\Delta X$ & counts & \multicolumn{1}{|p{1cm}|}{\centering corrected\\counts} & $dn/dX$ & \multicolumn{1}{|p{1.5cm}|}{\centering $\delta(dn/dX)$\\(+,-)}\\
		\hline
		5.169-5.679 & 5.51 & 38.88 & 6 & 6.77 & 0.17 & 0.10, 0.07\\
		5.679-6.381 & 5.89 & 38.90 & 13 & 14.09 & 0.36 & 0.13, 0.09\\
		\hline
    \end{tabular}
    
\end{table*}

\citet{cooper} (\citetalias{cooper} from here on) compared column densities of low and high ionization carbon (\ion{C}{ii}/\ion{C}{iv}) at $z>5$ using the spectra of 47 quasars from Magellan/FIRE and Keck/HIRES. \citetalias{cooper} found that at $z>5.7$, high ionization absorbers are very weak or undetected relative to low ionization absorbers. 

The E-XQR-30 has a higher spectral resolution when compared with \citetalias{cooper} (except for the one quasar from Keck/HIRES) and thus, it is worthwhile comparing how the improved data quality has affected the overall trend in \ion{C}{ii} evolution. There are 11 common quasar lines of sight with the E-XQR-30 sample and all systems reported in \citetalias{cooper} were recovered and 11 additional systems were also detected \citep{xqr30catalog}. Using 48 quasar spectra, \citetalias{cooper} detected 35 \ion{C}{ii} systems (16 among them are limits) at $z>5$ compared to the 46 \ion{C}{ii} (both proximate and non-proximate) detections in E-XQR-30. \citetalias{cooper} found that the highest redshift absorbers are more likely to be detected only in \ion{C}{ii} or \ion{Mg}{ii}. This has been further confirmed through the works of \citet{dodorico2010,dodorico2013} where the \ion{C}{iv} comoving line density and/or cosmic mass density decline with increasing redshift. The decrease in the densities of highly ionized carbon towards lower redshifts is also supported by the findings of \citet{becker2019} where they observe a decline in the equivalent width ratios of \ion{C}{iv} to \ion{O}{i} (detected in association with \ion{C}{ii} absorption lines) with redshift. Furthermore, \citet{Davies2023} (see Figure 7 in their paper) has illustrated the relative contributions of \ion{C}{II} and \ion{C}{iv} to the cosmic mass density evolution of carbon, where $\Omega_{\ion{C}{iv}}$ declines with redshift and $\Omega_\ion{C}{ii}$ increases across $5.2<z<6.4$.   

Other works on high redshift chemical enrichment such as \citet{becker2011} and \citet{bosman2017}, using their very small samples, found that the comoving line density evolution of low ionization absorbers such as \ion{C}{ii} at $5<z<7$ is similar to that of low ionization systems traced by damped Lyman alpha (DLA - log $N_\ion{H}{i}/cm^2\ge20.3$) and sub-DLA ($19.0<\text{log }N_\ion{H}{I}/cm^2<20.3$) across $3<z<5$, tentatively suggesting a constant evolution. Conversely, the $dn/dX$ calculated using the relatively larger sample with higher S/N data from E-XQR-30 show that the \ion{C}{ii} absorbers increase in density at higher redshifts. Also, cosmological hydrodynamic simulations such as \citet{finlator2015, keating2016} predicted a slow decrease in \ion{C}{ii} number density per absorption path contrary to the observed upturn for \ion{C}{ii} in this work. Considerable work has to be done in modelling the chemical enrichment of early Universe to match the observations. Furthermore, high resolution cosmological simulations on chemical enrichment of early Universe like \citep{oppenheimer2009} show that \ion{C}{ii} absorbers trace low mass galaxies at higher redshifts which are believed to be the major contributors of ionizing flux during EoR \citep{robertson, duffy, wise, finkelstein2019, Matthee, yeh}. 

\subsection{O {\scriptsize I} \texorpdfstring{$\lambda\ 1302$}{lone}\AA}
\label{subsec: OI}

\ion{O}{i} absorption line traces the neutral gas in the galaxy halos because its first ionization potential is very close to hydrogen and due to charge exchange $n(O^+)/n(O)\approx n(H^+)/n(H)$ over a wide variety of physical conditions \citep[][]{chambaud,osterbrock,becker2015,becker2019}. The E-XQR-30 detects 29 \ion{O}{i} systems out of which 10 are intervening absorbers (10,000 \kms proximity limit) in the primary sample across a redshift range of $5.3<z<6.4$ covering a path length of $\Delta X=46.359$. All \ion{O}{i} detections in the catalog have associated \ion{C}{ii} transitions \citep{xqr30catalog}.

As outlined in Section \ref{subsec: The low-ionization absorbers: dn/dX}, the absorbers are binned into two redshift bins covering equal $\Delta X$. After applying the 50\% completeness limit, the number of systems reduces to 8. The absorber counts in each redshift bin are completeness corrected and the resultant $dn/dX$ values are given in Table \ref{tab:3}.

\citet{becker2019} (from here on \citetalias{becker2019}) detected 57 intervening \ion{O}{i} systems using 199 quasar spectra from Keck/ESI and VLT/XSHOOTER across $3.2<z<6.5$ with a S/N ratio of 10 per 30\kms at a rest wavelength of 1285\AA. These absorbers are separated from the quasar emission redshift by a velocity separation of $>5000$ \kms. \citetalias{becker2019} reported a rapid increase in the comoving line density of \ion{O}{i} at $5.7<z<6.5$ covering a pathlength of $\Delta$X=66.3. This is $2.5^{+1.6}_{-0.8}$ times greater than the comoving line density across $4.9<z<5.7$. 

\begin{figure}
    \centering
    \includegraphics[width=\columnwidth]{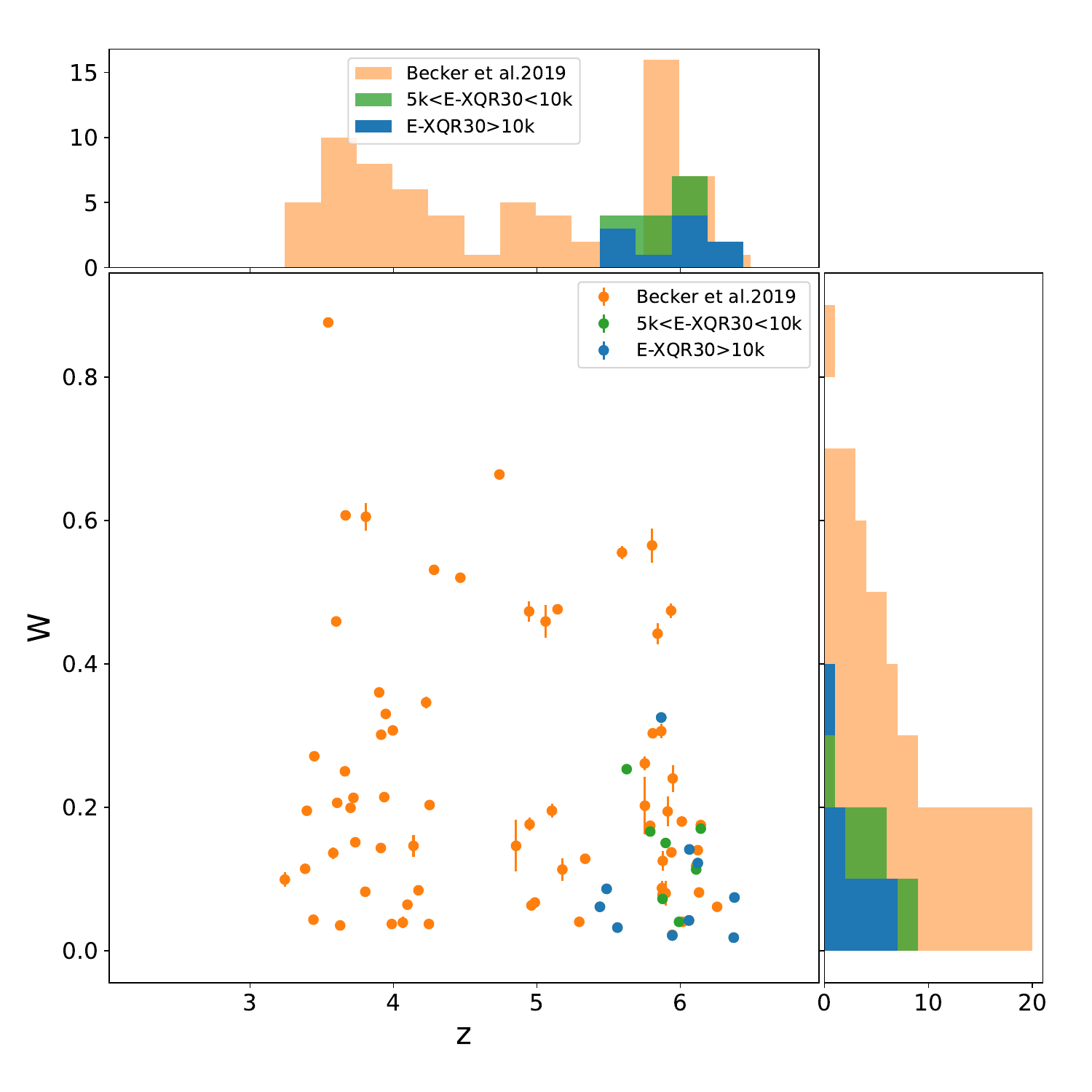}
    \caption{A scatter plot with histograms showing the sample distribution of \citetalias{becker2019} and E-XQR-30 intervening \ion{O}{i} absorbers. The blue colored scatter points represent E-XQR-30 data with a proximity limit of 10000\kms while the green coloured points show the E-XQR-30 \ion{O}{i} absorbers lying between 5000\kms and 10000\kms. The orange dots represent the \citetalias{becker2019} data which uses a proximity limit of 5000\kms. The top histogram gives the redshift distribution for \ion{O}{I} data sets from \citetalias{becker2019} and E-XQR-30 with different proximity limits. Similarly, the histogram on the right gives the rest-frame equivalent width distribution of the data in the scatter plot.The figure demonstrates that the E-XQR-30 sample consists of \ion{O}{i} absorbers with $W<0.4$\AA\ at $z>5.3$ compared to the \citetalias{becker2019} sample.}
    \label{fig:6}
\end{figure}

A scatter plot along with histograms that compare the distribution of intervening \ion{O}{i} absorbers along redshift and equivalent width between E-XQR-30 and \citetalias{becker2019} is shown in Figure \ref{fig:6}. The blue colour represents the data from E-XQR-30 using the proximity limit adopted in the catalog. The \ion{O}{i} systems from \citetalias{becker2019} that use 5000 \kms as proximity limit are colored orange. The errors in equivalent width are also shown in the scatter plot. The histogram on top shows the redshift distribution of both data sets. The E-XQR-30 has data only in the high redshift range, as the survey focuses only on quasars at $z>5.8$. The histogram on the right gives the equivalent width distribution of \ion{O}{i} data from both surveys. The E-XQR-30 data consist of \ion{O}{i} absorbers with $W\lesssim0.4$\AA\ where most of the absorber population from \citetalias{becker2019} lie in the scatter plot. However, the equivalent widths of data from \citetalias{becker2019} range between 0.02\AA$<W<$0.9\AA. 

To compare with \citetalias{becker2019}, the additional absorbers obtained by switching to a 5000 \kms\ proximity limit is shown in green in the scatter plot and histograms. The number of intervening absorbers increased from 10 to 17 and the absorption pathlength increased to $\Delta X=69.038$. Adopting the new proximity limit also improved the consistency of the $dn/dX$ results with \citetalias{becker2019}. 

\begin{figure}
    \centering
    \includegraphics[width=\columnwidth]{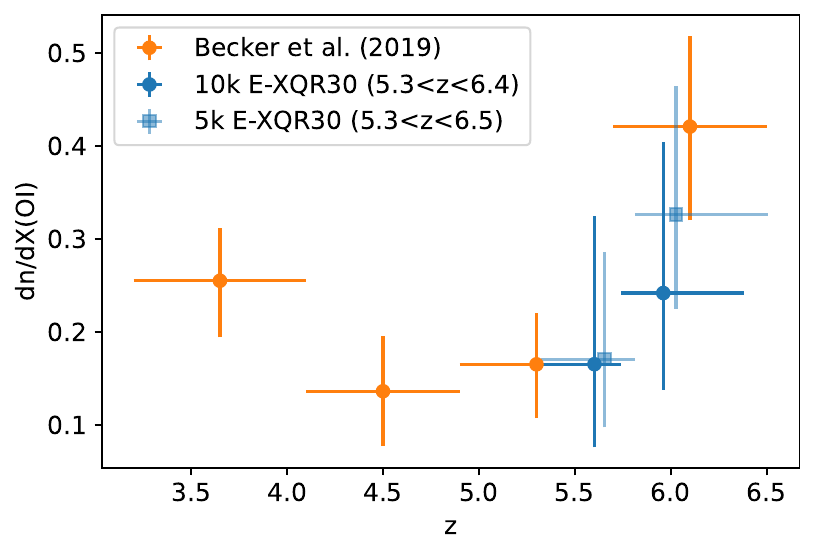}
    \caption{The comoving line density of \ion{O}{i} using E-XQR-30 in comparison to the work of \citetalias{becker2019}. Only \ion{O}{i} absorbers with $W>0.03$\AA\ are used in the analysis. The dark blue circles show the $dn/dX$ calculated for \ion{O}{i} across $5.3<z<6.4$ by applying a proximity limit of 10000\kms while the light blue squares are the $dn/dX$ values for \ion{O}{i} across $5.3<z<6.5$ using a proximity range of 5000 \kms from E-XQR-30. The orange points show the $dn/dX$ values in \citetalias{becker2019} which used 5000\kms as the proximity limit. The figure shows an upturn in the comoving line density of \ion{O}{i} at $z>5.3$ which is consistent with \citetalias{becker2019} results.}
    \label{fig:7}
\end{figure}

Figure \ref{fig:7} depicts the redshift evolution of comoving line density of \ion{O}{i} absorbers from E-XQR-30 with different proximity limits together with the results from \citetalias{becker2019}. The dark blue points are the $dn/dX$ for data from E-XQR-30 using a proximity limit of 10000 \kms and the light blue squares are those with a proximity limit of 5000 \kms. The results from the work of \citetalias{becker2019} is shown in orange dots. It is interesting to observe the upturn at $z\gtrsim5.7$ even with the small number of \ion{O}{i} absorbers from our survey. Cosmic variance may offer an explanation to the minor discrepancies in the $dn/dX$ values between our work and \citetalias{becker2019}. However, the values agree with \citetalias{becker2019} within the $1\sigma$ error bars. 
\begin{table*}
    \centering
    \caption{The $dn/dX$ values of \ion{O}{i} using different proximity limits. The proximity limits, redshift ranges, pathlength weighted mean redshift, absorption pathlength, raw counts and completeness corrected counts, associated errors with $dn/dX$ are also given in the table. }
    \label{tab:3}
    \begin{tabular}{|l|c|c|c|c|c|c|r|}
        \hline
		\multicolumn{1}{|p{1cm}|}{\centering proximity\\ (\kms)} & $z$ range & \textlangle $z$\textrangle & $\Delta X$ & counts & \multicolumn{1}{|p{1cm}|}{\centering corrected\\counts} & $dn/dX$ & \multicolumn{1}{|p{1.5cm}|}{\centering $\delta(dn/dX)$\\(+,-)}\\
		\hline
		\multirow{2}{2em}{10000} & 5.322-5.742 & 5.60 & 23.17 & 3 & 3.83 & 0.17 & 0.16, 0.09\\
            & 5.742-6.381 & 5.96 & 23.19 & 5 & 5.61 & 0.24 & 0.16,0.10\\
            \hline
		\multirow{2}{2em}{5000} & 5.322-5.813 & 5.65 & 34.51 & 5 & 5.89 & 0.17 & 0.12, 0.07\\
		& 5.813-6.505 & 6.03 & 34.53 & 10 & 11.26 & 0.33 & 0.14, 0.10\\
		\hline
    \end{tabular}
\end{table*}

Only 21 quasars in their sample are included in our survey and \citet{xqr30catalog} has successfully recovered all \ion{O}{i} systems reported in \citetalias{becker2019} along with the detection of one additional system. Sondini et. al. (2023, in prep) analyse the E-XQR-30 data and report two additional \ion{O}{I} absorbers that fall strictly outside the Lyman $\alpha$ forest cutoff adopted by the \citet{xqr30catalog} catalog: PSO J060+24 at $z=5.6993$ and SDSSJ0100+2802 at $z=5.7974$. There is also a chance of clustering of \ion{O}{i} absorbers at high redshift $z\sim6$ as reported in \citetalias{becker2019} due to the fluctuations in UV background towards the tail end of EoR \citep[e.g.,][]{D'Aloisio2018, kulkarni2019}. Completeness variations are explored in Appendix \ref{new OI dn/dX}.

\subsection{Comparison with JWST/NIRSpec results on metal enrichment}
\label{JWST comparison}

Recently, \citet{Christensen2023} observed four quasar sightlines at $z>6.5$ using NIRSpec on JWST (James Webb Space Telescope) and studied the evolution of metal enrichment at $2.42<z<7.48$. Their quasar spectra cover a wider spectral range and have high S/N of 50-200 but with lower spectral resolution of $R\simeq2700$ compared with E-XQR-30. Using the four quasar spectra, they detected 61 systems at $2.42<z<7.48$ and calculated the comoving line density of these systems including the ions that we studied in this work. The one quasar common in both works is J0439+1634 at $z=6.52$. \citet{xqr30catalog} detected 43 absorption systems from the spectrum, while \citet{Christensen2023} detected only 17 systems including an additional \ion{Mg}{ii} doublet at $z=6.208$. The non-detected systems include several weak \ion{Mg}{ii} absorbers, weak \ion{C}{iv} doublets and \ion{Si}{iv} doublets at $2.3<z<5.3$ \citep{Christensen2023} emphasising the critical need for median resolution spectroscopy in the JWST-era.

The proximity limit adopted by \citet{Christensen2023} is 3000 \kms although J0439+1634 is affected by BAL regions to larger velocities of $\sim10,000$ \kms. They examined the redshift evolution of low ionization absorbers using 6 \ion{O}{i} (two of them are new detections at $z>7$), 8 \ion{C}{ii} intervening systems at $6.4<z<7.4$ and observed an upturn for \ion{O}{i} and \ion{C}{ii} at $z>6$. They also studied the evolution of 48 \ion{Mg}{ii} (one of them is a new detection at $z\sim7.443$) intervening systems with $W>0.3$\AA\ at $2.4<z<7.45$ and observed a constant evolution for \ion{Mg}{ii}. However, $dn/dX$ values across redshift in our work using a large sample of \ion{Mg}{ii} show that they decline with redshift at least until $z\sim5$. Their work also illustrates a decline in strong \ion{Mg}{ii} comoving line density evolution. Overall, the results from JWST / NIRSpec are consistent with the results from this work for \ion{O}{i}, \ion{C}{ii} and strong \ion{Mg}{ii}.

\section{Discussion}
\label{sec:Discussion}

\subsection{The nature of weak Mg {\scriptsize II} absorbers at high redshift}
\label{subsec:weak MgII absorbers at high z}

This work presents for the first time a significant population of 131 intervening weak \ion{Mg}{ii} absorbers sufficient to produce robust statistical results in their evolution. There are several low redshift ($z<2$) studies in the literature finding that weak absorbers are distributed along the major axis of face on, blue galaxies \citep{kapzak2012, nielsen2015} tracing the infalling gas into the CGM and they have little correlation between the absorber equivalent width and the galaxy colour \citep{Chen2010, kapzak2011}. Furthermore, \citet{Churchill2005} and \citet{Chen2010} have showed that weak \ion{Mg}{ii} absorbers do not necessarily trace low surface brightness galaxies by studying galaxy absorber pairs at intermediate and low redshifts respectively. Recent high redshift metal absorber surveys reported detections of weak \ion{Mg}{ii} absorbers, however, there were not enough sightlines or sensitivity to produce a large population of weak absorbers sufficient to measure their cosmic evolution. For example, \citet{chen} reported 59 detections of weak \ion{Mg}{ii} absorbers, however, their overall completeness was not enough to produce robust statistics on them. Similarly, works by \citet{codoreanu} and \citet{bosman2017} also presented weak \ion{Mg}{ii} detections, but they were also limited in their sample size (10 systems at $2<z<5$ and 3 systems at $z>5.5$ respectively). 

In this work, weak \ion{Mg}{ii} absorbers account for 47\% of the total completeness corrected number of intervening \ion{Mg}{ii} absorbers in E-XQR-30. It has been observed that the evolution of weak \ion{Mg}{ii} absorbers remain constant - within a factor of 2 - in a redshift range of $1.9<z<6.4$ from Figure \ref{fig:2}. 

Much literature exists on the association of weak \ion{Mg}{ii} systems and the relative number density of LLSs, the role of SFR and metallicity trends, dwarf galaxies and the UVB at $z < 2$. With the benefit of a long lever arm from z=1.9 to 6.4, these hypotheses can be tested. At lower redshifts, $0.4\le z\le1.4$, it is assumed that the weak \ion{Mg}{ii} absorbers arise in the sub-LLS environments because the weak absorbers outnumber the LLS absorbers in comoving line density \citep{churchill}. \citet{crighton}, using a data set of 153 optical quasar spectra from the Giant Gemini GMOS survey ($3.5 < z < 5.4$) and complementary literature data to calculate the incidence of LLS absorbers per unit absorption pathlength ($dn/dX$), which increases steeply from redshift 0 through to 5.4. Since the $dn/dX$ of weak \ion{Mg}{ii} absorbers is flat, swapping from being more numerous than LLSs at $z<2$ to three times rarer at $z>4$, is only possible if the covering fraction varies by the same amount. \citet{dutta2020} find that the covering fraction of weak \ion{Mg}{II} systems does not change from low redshift out to $z\sim1.5$. Nevertheless, due to the metallicity and the ionization effects at high z, the neutral hydrogen column density of systems detected as weak \ion{Mg}{ii} increases and consequently, LLSs could be associated with weak \ion{Mg}{ii} absorbers \citep{steidel1992, churchill, Rigby2002}. Our results support this picture with a factor of 3 drop in the covering fraction out to redshift $\sim6$. 

Along the same lines, we can now examine the association of weak \ion{Mg}{ii} absorbers at $0 < z < 2.4$ and the star formation history of dwarf galaxies \citep{churchill, Lynch2006, lynch, 2008Narayanan}, which both peak at $z\sim1$ and drop at $z\sim2$. Knowing that $dn/dX$ continues at a constant level to $z\sim6$, indicates this hypothesis cannot continue to high redshift. As mentioned in Section \ref{subsec: MgII}, \citetalias{chen} were able to reconcile the \ion{Mg}{ii} $dn/dX$ evolution to $z\sim6$ by allowing the associated galaxy mass to decrease with increasing redshift. Whatever the mechanisms that give rise to weak \ion{Mg}{ii}, it should be able to replenish them in such a way that the comoving line density of these low equivalent width absorbers do not change considerably across cosmic time.

\begin{figure*}
     \centering
     \begin{subfigure}[b]{0.4\textwidth}
         \centering
         \includegraphics[width=\textwidth]{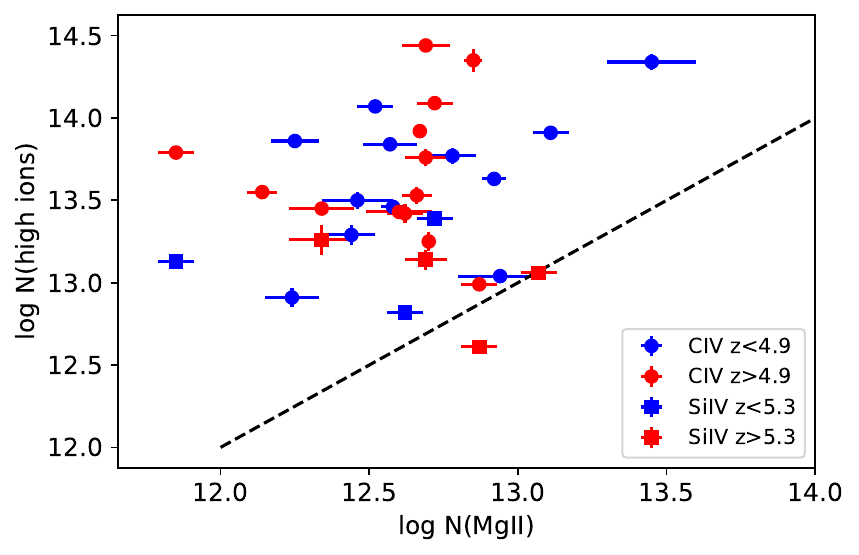}
         \caption{}
         \label{fig:weak_MgII_high_ions_logN}
     \end{subfigure}
     \hfill
     \begin{subfigure}[b]{0.4\textwidth}
         \centering
         \includegraphics[width=\textwidth]{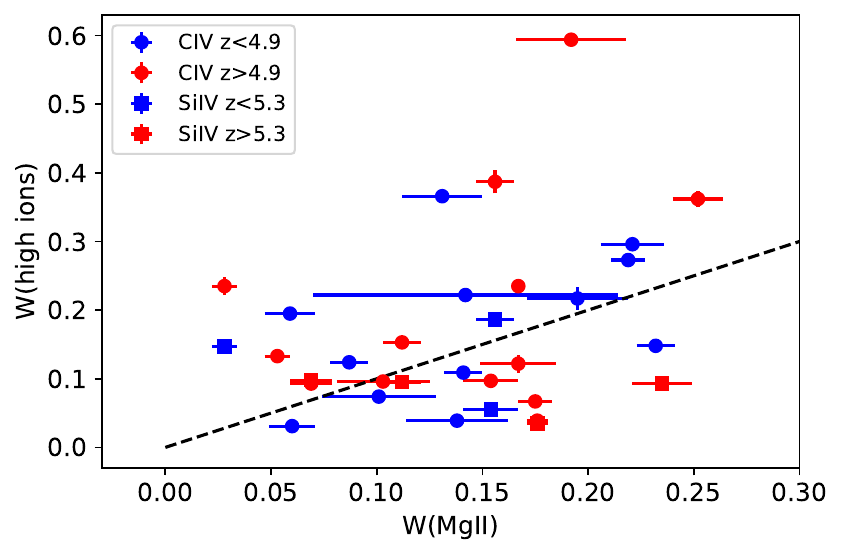}
         \caption{}
         \label{fig:weak_MgII_high_ions_W}
     \end{subfigure}
     \caption{The left panel shows the column densities of high ionization absorbers, \ion{C}{iv} and \ion{Si}{iv} versus the column densities of weak \ion{Mg}{ii} absorbers. The right panel shows the rest-frame equivalent widths of the \ion{C}{iv} and \ion{Si}{iv} absorbers against the equivalent widths of weak \ion{Mg}{ii} absorbers. Both the associated \ion{C}{iv} and \ion{Si}{iv} are divided based on the median value of the redshift for the sample. The dots represent the \ion{C}{iv} sample; blue indicates the lower redshift and red indicates the higher redshift samples. The squares stand for the \ion{Si}{iv} sample; blue showing the lower redshift and red showing the high redshift sample. The dashed line represents the 1:1 relation between the respective parameters of the high ionization absorbers and weak \ion{Mg}{ii}. The figure depicts that weak \ion{Mg}{ii} systems have higher column densities of \ion{C}{iv} and \ion{Si}{iv} and both weak \ion{Mg}{ii} and high ionization absorbers have comparable equivalent widths.}
        \label{fig:weak_MgII_high_ions}
\end{figure*}

The weak \ion{Mg}{ii} absorbers are also associated with high ionization absorbers such as \ion{C}{iv} and \ion{Si}{iv} indicating a multi-ionization phase for these weak absorbers. Among 58 intervening weak \ion{Mg}{ii} absorbers from primary sample across the redshift range where \ion{C}{iv} and \ion{Si}{iv} can be detected, 25 of them are associated with \ion{C}{iv} at $4.3\le z\le 6.3$ and 7 of them are associated with \ion{Si}{iv} at $4.9\le z \le 6.3$. We have not used the 50\% completeness cut here to find out the fraction of \ion{Mg}{ii} absorbers with high ionisation species. This constitutes 43\% of weak \ion{Mg}{ii} absorbers with \ion{C}{iv} across $\Delta$X=211.8 and 25\% of weak \ion{Mg}{ii} with \ion{Si}{iv} across $\Delta$X=120.6. Among the weak \ion{Mg}{ii} absorbers associated with high ionization absorbers, 6 of them  have both \ion{C}{iv} and \ion{Si}{iv} detections. Most of these weak \ion{Mg}{ii} absorbers are single component systems at the resolution of XSHOOTER. Both the \ion{C}{iv} and the \ion{Si}{iv} absorbers are detected in weak \ion{Mg}{ii} systems up to a redshift of $z\sim5.9$ after which no \ion{C}{iv} or \ion{Si}{iv} associations are found. This might be because of the decrease in high ionization absorber comoving line density \citep{cooper,d'odorico2022, Davies2023} at higher redshifts due to lower metal abundance and a softer ionizing background \citep{cooper,finlator2016}. We also examined whether the absence of weak \ion{Mg}{ii} systems with \ion{C}{iv} and \ion{Si}{iv} at $z>5.9$ is due to the lower S/N of the quasar spectra towards these redshifts and found that the typical S/N of the spectra is roughly equivalent at both $4.3<z<5.9$ and $5.9<z<6.3$. The relation between the column densities of the high ionization absorbers, \ion{C}{iv} and \ion{Si}{iv}, against the column densities of weak \ion{Mg}{ii} absorbers is shown in Figure \ref{fig:weak_MgII_high_ions_logN}. Similarly, the equivalent widths of \ion{C}{iv} and \ion{Si}{iv} are plotted against the equivalent widths of weak \ion{Mg}{ii} in Figure \ref{fig:weak_MgII_high_ions_W}. In both figures, the dots represent \ion{C}{iv} detections associated with weak \ion{Mg}{ii} and the squares represent \ion{Si}{IV} associated with weak \ion{Mg}{ii}. Both the samples are divided based on their median redshift and the low redshift sample is depicted using blue colour and the high z sample is shown using red colour. A dashed line representing the 1:1 relation between the column densities (equivalent widths) of high ionization absorbers and the weak \ion{Mg}{ii} absorbers is marked on both the figures. It can be observed from the left panel in Figure \ref{fig:weak_MgII_high_ions} that half of the weak \ion{Mg}{ii} absorbers in the E-XQR-30 sample are rich in triply ionized carbon and silicon because most of the points lie on the upper region of the 1:1 line. The right panel in Figure \ref{fig:weak_MgII_high_ions} shows that the strengths of the absorbers are comparable between the high ionization absorbers and weak \ion{Mg}{ii} at the probed redshift ranges. These weak systems are exposed to ionizing radiation strong enough to ionize the associated carbon and silicon in them to their triply ionized state. Moreover, these high ionisation absorbers detected in the weak \ion{Mg}{ii} systems are indicative of metal-enriched material in the CGM, rather than the metal-poor infalling gas from the IGM. Whether the high-$z$ weak(strong) \ion{Mg}{ii} systems follow the same origin pattern of corotating/infalling (outflowing) gas in the CGM as their low-$z$ counterparts will need to wait for detailed absorber-galaxy pair kinematic analysis. We also looked at the redshift dependence of the ratios of column densities and equivalent widths between weak \ion{Mg}{ii} and highly ionized absorbers and no clear trend is observed in these systems with respect to redshift. 

\begin{figure}
    \centering
    \includegraphics[width=\columnwidth]{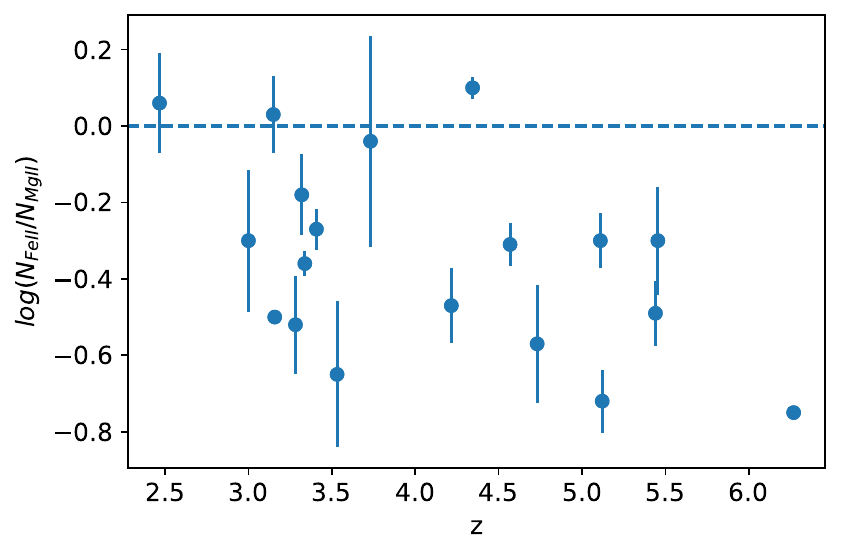}
    \caption{The ratio of column densities of \ion{Mg}{ii} and \ion{Fe}{ii} across redshift. The errors are the 1 $\sigma$ statistical errors in the column densities of bith the ions. Out of the 21 weak \ion{Mg}{ii} absorbers, three are iron rich systems (above the dashed horizontal line) which disappear at $z>4.5$.}
    \label{fig:weak MgII FeII}
\end{figure}

Studying iron abundance ratios in weak \ion{Mg}{II} absorbers can provide further clues to their origin. 21 weak \ion{Mg}{ii} absorbers at $2.5<z<6.3$ are associated with \ion{Fe}{ii} whose $\text{log}\ (N_{\ion{Fe}{II}}/N_{\ion{Mg}{II}})$ values range between -0.75 to 0.1. Among these \ion{Fe}{ii}-\ion{Mg}{ii} associations, three are found to be iron-rich ($\text{log}\ (N_{\ion{Fe}{ii}}/N_{\ion{Mg}{ii}})\ge0$) as shown in Figure \ref{fig:weak MgII FeII}. There appears to be an anti-correlation between the column density ratios of the absorbers and the redshift. The iron-rich systems above the dashed horizontal line seem to disappear at $z > 4.5$, but this might be due to the relatively small number of absorbers detected at higher redshifts. A K-S test of the systems at $z<4.5$ and $z>4.5$ yielded a p value of 0.15 showing that $z > 4.5$ and $z < 4.5$ \ion{Mg}{II} absorbers most probably arise from the same parent population. A similar anti-correlation has been reported by \citet{2008Narayanan} in a study of weak \ion{Mg}{ii} absorbers at $0.4<z<2.4$. A detailed study on the association of \ion{Mg}{ii} and \ion{Fe}{ii} might give us some hints about their absence. However, according to previous low redshift studies on the association of weak \ion{Mg}{ii} absorbers and \ion{Fe}{ii} \citep{Rigby2002,2008Narayanan}, these iron-rich weak \ion{Mg}{ii} absorbers require enrichment from Type Ia supernovae (SNe) while the iron-poor systems require alpha enhancement from external enrichment such as bubbles and superwinds of massive galaxies or trapping of ejecta from local SNe \citep{Rigby2002}. The observed increase in alpha enhancement of weak \ion{Mg}{ii} systems with redshift shows that the most of the early galaxies traced by these systems are producing more alpha elements through core-collapse supernovae of massive stars in short timescales \citep{thomas1999, thomas2010, johansson, conroy, segers}. Nevertheless, galaxies associated with the iron-rich systems in our sample likely result from stellar products over longer timescales, possibly in the inner halos of massive quiescent galaxies \citet{zahedy2017}.

\subsection{Strong Mg {\scriptsize II} absorbers and star formation history}
\label{subsec:strong MgII}

There have been a handful of previous works showing the potential of strong \ion{Mg}{ii} absorbers in tracing the star formation history (SFH) of the Universe. Earlier works of strong \ion{Mg}{ii} have found out that there are correlations between the equivalent width of the absorber and the blue host galaxy colour \citep{zibetti2007, lundgren2009, noterdaeme2010, bordoloi2011, nestor2011}. Strong \ion{Mg}{ii} absorbers are also observed to be associated with star-forming galaxies from their spectroscopic observations \citep{weiner2009, rubin2010}. Studies by \citet{kapzak2012} and \citet{nielsen2015} on the azimuthal angle dependence of \ion{Mg}{ii} have shown that strong \ion{Mg}{ii} absorbers are found along the minor axis of face-on red galaxies tracing outflows from galaxies. On the contrary, there also exist studies like \citet{bouche2016, zabl2019} showing that \ion{Mg}{II} absorbers are aligned with the galactic discs but at a lower incidence rate, tracing the inflows. \citet{menard2011} formulated a scaling relation between the median \ion{O}{ii} luminosity surface density and the equivalent width of the \ion{Mg}{ii} absorbers at $0.4<z<1.3$ and showed that strong \ion{Mg}{ii} absorbers are powerful tracers of star formation independent of redshift and not remarkably affected by dust extinction. Near-IR spectroscopic surveys such as \citet{matejek} and \citet{chen} showed that the decline in cosmic SFR  after the reaching the peak at $z\sim2$, is reflected in the observed decline in the comoving line density of strong \ion{Mg}{ii} at $z>2$. They used the scaling relation from \citet{menard2011} to estimate the SFR from \ion{Mg}{ii} equivalent widths and found them to be in agreement with the observed SFR calculation from Hubble Space Telescope \citep{bouwens2010, bouwens2011}. 
Moreover, many previous works on \ion{Mg}{ii} absorbers such as \citet{prochter, zibetti2007, lundgren2009, noterdaeme2010,bordoloi2011,nestor2011} conclude that strong \ion{Mg}{ii} absorbers arise in galactic outflows.  

\begin{figure}
    \centering
    \includegraphics[width=\columnwidth]{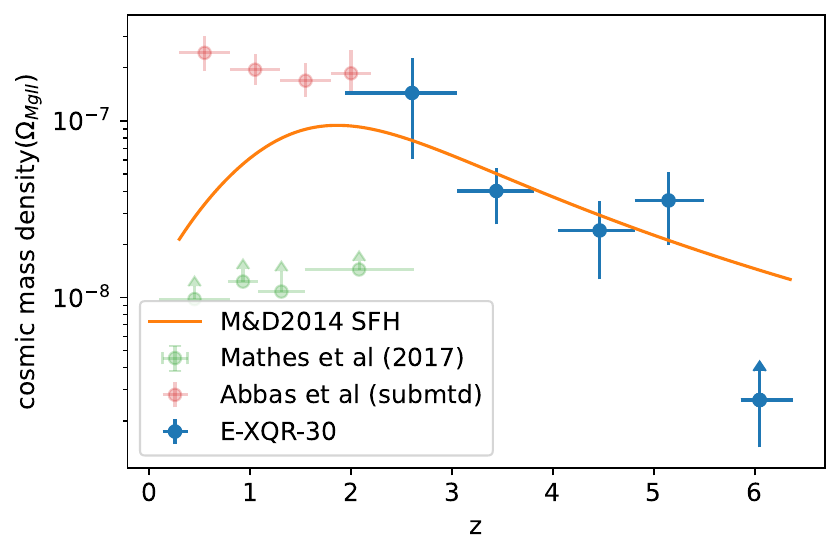}
    \caption{The figure shows the evolution of cosmic mass density of \ion{Mg}{ii} dominated by strong \ion{Mg}{ii} at $1.9<z<6.4$. The blue points represent the $\Omega_{\ion{Mg}{ii}}$ and the orange line indicates the cosmic SFH from \citet{madau-dickinson} normalized to the observations. $\Omega_\ion{Mg}{ii}$ measurements at $0<z<2$ are also shown for comparison. It can be observed that $\Omega_{\ion{Mg}{ii}}$ declines with redshift at $z>3$ with little to no evolution towards $z<3$ which is consistent with the global SFH trend. Since cosmic mass density is a mass-weighted statistic, the sharp decline in $\Omega_{\ion{Mg}{ii}}$ at $z>5.5$ is attributed to zero strong \ion{Mg}{II} systems detected in the final redshift bin, as explored in Figure \ref{fig:W norm hist}.}
    \label{fig:omega mgii}
\end{figure}

In the E-XQR-30 sample of  280 intervening \ion{Mg}{ii} absorbers, 53 absorbers have $W>1.0$\AA\  and due to the improved sensitivity of X-SHOOTER, we were able to measure the column densities of these absorbers. Consequently, the cosmic mass density ($\Omega_{\ion{Mg}{ii}}$ - mass of the absorber per unit comoving Mpc/critical density of the Universe) of the strong absorbers can be calculated using the following equation based on the approximation of \citet{SL96} \begin{equation}\label{eq:omega mgii}
    \Omega_\text{\ion{Mg}{ii}} \simeq \frac{H_0 m_{\text{Mg}}}{c\rho_{crit}\Delta X}\Sigma_i \frac{N_{\text{total},i}}{C(z,\text{log}N_i)}
\end{equation} where $H_0$ is the Hubble constant, $m_{\text{Mg}}$ is the atomic mass of magnesium, $\rho_{crit}$ is the critical density of the Universe, $N_{\text{total}, i}$ is the column density of the absorbers in the \textit{i}th log $N$ bin and $C(z,\text{log}N_i)$ is the completeness correction. The errors associated with $\Omega_\ion{Mg}{ii}$ is computed using the equation below from \citet{SL96} \begin{equation}
    \bigg(\frac{\delta\Omega_\ion{Mg}{ii}}{\Omega_\ion{Mg}{ii}}\bigg)^2 = \frac{\Sigma(N^2)}{(\Sigma N)^2}.
    \end{equation} In the literature, $\Omega_\ion{Mg}{II}$ has been compared with the cosmic SFH evolution, since both are functions of volume density. However, our long lever arm from $2<z<6$ highlights the differences between the two measures: the turnover in cosmic SFR at $z \sim 2$ is contrasted with the cosmic mass density of the \ion{Mg}{II} ion, a fraction of total metal density, which is monotonic over time. Our results on the cosmic mass density of \ion{Mg}{ii} at $2<z<5.5$ agree well with the trend in the global star formation rate density from \citet{madau-dickinson} that has been normalised to the $\Omega_{\ion{Mg}{ii}}$ measurements, as shown in Figure \ref{fig:omega mgii}. The values are tabulated in Table \ref{tab:1}. The blue points refer to the cosmic mass density measurements from this work, and the orange curve represents the best fitting function for the global SFH from \citet{madau-dickinson} at $0.3<z<6.4$. The light green points are $\Omega_\ion{Mg}{ii}$ from \citet{mathes2018} and the light red points are from \citet{abbas} for $z<2$. It can be observed that the mass density shows little to no-evolution for $z\le2$. As can be seen in Figure \ref{fig:omega mgii}, the mass densities from \citet{mathes2018} is an order of magnitude lower than those from \citet{abbas}. This is because \citet{mathes2018} used the apparent optical depth method to measure the column densities of \ion{Mg}{ii} systems, which provides only lower limits for unresolved saturation and therefore, should be viewed as lower limits \citep{abbas}. Overall, the cosmic mass density of \ion{Mg}{ii} shows little evolution from redshift 0 to 3,  followed by a drop in $\Omega_\ion{Mg}{II}$ at $z>3$ which is consistent to the expected trend in the metal content evolution of the Universe from the global SFH \cite[see][Figure 14]{madau-dickinson}. As expected, there is a lag between SFR and metals appearing in the CGM (see Figure 6 in \citet{Davies2023} for examples of the predicted rate of change in metal content flowing through to the measurement of $\Omega_\ion{C}{IV}$). The peak in the metal mass density is anticipated to be at lower redshifts than the peak in the global SFH and therefore, the slight rise in $\Omega_\ion{Mg}{II}$ after the cosmic noon is consistent with the cosmic SFH trend. Nevertheless, \citet{menard2011} have shown that strong \ion{Mg}{II} trace a substantial fraction of global star formation at redshifts $z<2$ based on an SDSS survey. Though $\Omega_\ion{Mg}{II}$ from this work hints at a sudden drop in mass density at $z>5.5$ due to the absence of strong \ion{Mg}{II}, the small $\Delta X$ of the final redshift bin means that the null detection of strong absorbers is not statistically significant. A more detailed analysis of the relation between SFH and metal absorbers will be discussed in an upcoming work on galaxy-absorber pairs.

\begin{figure}
    \centering
    \includegraphics[width=\columnwidth]{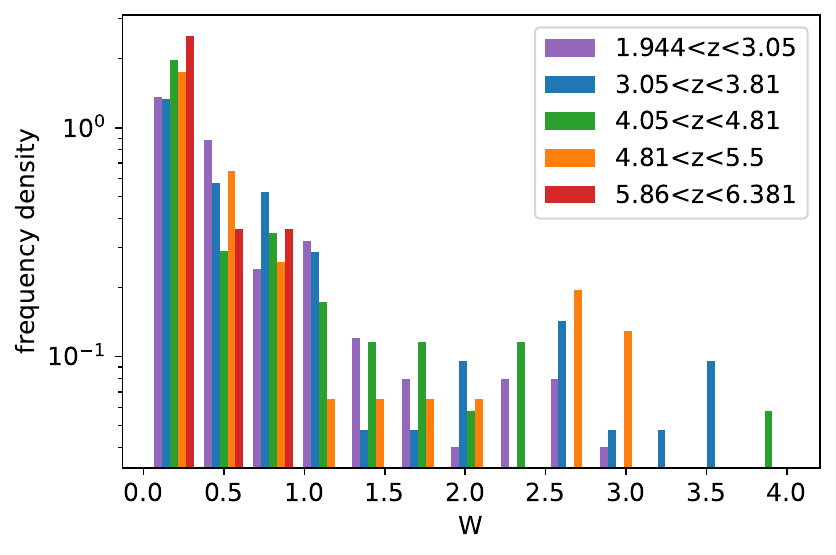}
    \caption{The rest-frame equivalent width distribution for \ion{Mg}{ii} that is colour coded based on the redshift interval to which the absorbers belong. The $W$ bins have a binwidth of $\sim0.3$\AA\ where each bin in each redshift interval has a binsize of $\sim0.05$\AA. The histogram is normalised to the total absorber counts in such a way that the area under the histogram equals to 1. The absorbers at all redshifts peak at $W<0.5\text{\AA}$ showing that majority of the absorbers in the E-XQR-30 sample are weak. As we progress to larger equivalent widths of $W>1.0\text{\AA}$, absorbers at $z>5.86$ disappear indicating an absence of high z strong \ion{Mg}{II} systems.}
    \label{fig:W norm hist}
\end{figure}

Although strong \ion{Mg}{ii} absorbers are subdominant at all redshifts compared to weak absorbers, their complete absence at $z>5.86$ in the E-XQR-30 sample warrants investigation. On one hand, \citetalias{chen} detected 1 intervening strong \ion{Mg}{ii} absorber at $z>5.9$ in one (J1048-0109 at $z\sim6.64$; not included in the E-XQR-30 sample) among 28 quasar spectra at these redshift ranges. On the other hand, \citet{keating2016} struggled to reproduce strong \ion{Mg}{ii} absorbers at z=6 in their simulated spectra. However, there are no strong absorbers at $z>5.8$ in our sample as can be found in the histogram showing equivalent widths of \ion{Mg}{ii} in different redshift intervals in Figure \ref{fig:W norm hist}. 
The same redshift intervals used in the $dn/dX$ analysis of the \ion{Mg}{ii} absorbers are applied here in the histogram where the binsize for $W$ is $\sim0.3$\AA\ that are sub-divided into bins of size $\sim0.05$\AA\ for each redshift interval. However, this is not sufficient evidence to conclude that strong \ion{Mg}{ii} absorbers are completely absent in the early Universe. There exist both possibilities of strong absorbers being either rare or less detected at higher redshifts. The initial step to investigate the absence of strong absorbers at high redshift is to check whether it is feasible to find strong absorbers at $z>5.86$ assuming that the equivalent width distribution is constant at all redshifts.
The analysis yielded an expectation value of detecting 3 strong absorbers at $z>5.86$. 
Using a Poisson distribution with the mean value as the expected number of strong absorbers at $z>5.86$, the probability of detecting no strong absorbers at $z>5.86$ is calculated, giving a value of 7\%. This points out that the non-detection of strong absorbers at $z>5.86$ is not statistically significant.  

Other properties of the strong \ion{Mg}{ii} absorbers were investigated, including their association with high ionization absorbers and \ion{Fe}{ii}. The medium and strong \ion{Mg}{ii} absorbers have fewer detections of associated \ion{C}{iv} and \ion{Si}{IV} in comparison with the weak absorbers because most of the strong \ion{Mg}{ii} absorbers in the E-XQR-30 sample are present at $z<4$ (see Table \ref{tab:3}) at which \ion{C}{IV} and \ion{Si}{IV} are inaccessible. However, the fractions of strong \ion{Mg}{II} associated with \ion{C}{IV} and \ion{Si}{IV} are higher (76\% and 40\% respectively) than weak absorbers; possibly indicating that these dense systems located closer to the galaxies (see Section \ref{subsec:chemical enrichment or UV}) are probably fragments from highly energetic galactic outflows. Among the few detections, most of them have higher column densities of \ion{Mg}{ii} relative to high ionization absorbers. Apart from being \ion{Mg}{ii} enhanced, the strong absorbers outnumber the weak \ion{Mg}{II} absorbers for their associations with \ion{Fe}{II} and have more iron-rich systems than weak absorbers. The increased number of iron rich systems among strong \ion{Mg}{ii} absorbers can imply their origins in Type Ia SNe driven galactic superwinds. 

On a side note, medium \ion{Mg}{ii} absorbers (0.3\AA$<W<$1.0\AA) might be tracing the same objects as those of the weak \ion{Mg}{ii} absorbers at high redshifts due to the similar trend in their comoving line density as observed in Figure \ref{fig:2}. But, certain aspects, such as relatively larger number of medium systems with high column density ratios of \ion{Mg}{ii} over high ionization absorbers and higher number of iron rich systems, when compared to the weak absorbers, tend to place the medium absorbers somewhere in between the weak and strong absorber properties. 

\subsection{Low ionization Absorbers Evolution: Chemical Enrichment or ionizing Background?}
\label{subsec:chemical enrichment or UV}

The comoving line density evolution of low ionization absorbers (hereafter LIAs, a term used by \citetalias{cooper}) from E-XQR-30 metal absorber catalog is a useful probe in understanding the chemical enrichment of the Universe and the strength of ionizing radiation after the EoR. The general trend observed in the evolution of these absorbers is that they increase towards high redshifts. The $dn/dX$ of \ion{C}{II} (Figure \ref{fig:4}) and \ion{O}{i} (Figure \ref{fig:7}) shows an upturn at $z>5.7$ . The \ion{Mg}{ii} absorbers, which are observed to arise in different neutral hydrogen column densities from $15.5<\text{log} N(\ion{H}{i})<20.5\ \text{cm}^{-2}$ \citep{bergeron,steidel1992,churchill2000}, trace a range of ionization states whose collective number densities decline with increase in redshift (Figure \ref{fig:1}). 

\begin{figure}
    \centering
    \includegraphics[width=\columnwidth]{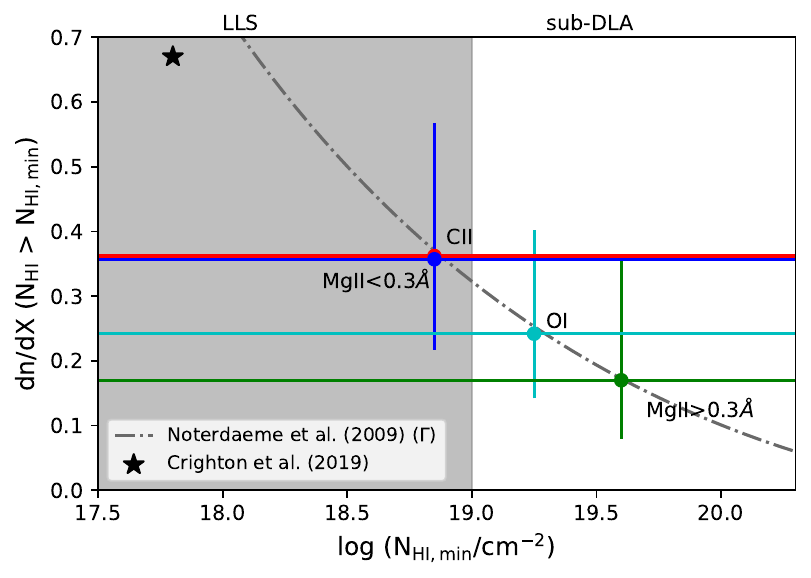}
    \caption{The DLA incidence rate as a function of the lowest \ion{H}{i} column density using the \ion{H}{i} column density distribution functions at $2<z<5$ in order to test the \citetalias{cooper} hypothesis of LIAs as analogues of metal poor DLAs and sub-DLAs. The gray dot-dashed curve shows the DLA distribution from previous literature as indicated in the plot. But due to the flattening of f(N$_{\ion{H}{i}}$,X) at the Lyman limit, the extrapolation of a single power law could be associated with large uncertainties towards $N_{\ion{H}{i}}<10^{19}\text{cm}^{-2}$. The black star marks the $dn/dX$ of LLS at z=5.05 from \citet{crighton}. The solid horizontal lines mark the incidence rate of the LIAs at $z>5.7$, different ion represented by a different colour. The grey shaded region represents the LLS column densities while the unshaded region shows the sub-DLA region. It can be seen that the comoving line density of each metal ion corresponds to a different absorber class based on the minimum HI column density. The figure shows that \ion{O}{I} and stronger \ion{Mg}{II} absorbers have a similar comoving line density to the sub-DLAs in \citet{omeara2007}. However, the higher incidence rates of \ion{C}{II} and weak \ion{Mg}{II} absorbers correspond to \ion{H}{I} absorbers that sit between the LLS and sub-DLA systems.}
    \label{fig:dla and sub-dla}
\end{figure}

One of the most intriguing questions about the LIAs at high redshift is about the properties of these systems in the early Universe. There have been many works trying to identify what these absorbers are associated with and how different they are from the low redshift absorber populations. Following \citetalias{cooper}, we use \ion{H}{i} column density distribution function ($f(N_{\ion{H}{i}},X)=\frac{\partial^2n}{\partial n \partial X}$) from the literature for DLAs at $2.0<z<5.0$ in \citet{noterdaeme2009} to compare with the $dn/dX$ of LIAs. The gamma function fit from \citet{noterdaeme2009} has been used here because this work uses a large sample of DLAs compared to other DLA studies in the literature and moreover, the extrapolation of the function reproduces the sub-DLA distribution. Figure \ref{fig:dla and sub-dla} depicts the incidence rate of DLAs and sub-DLAs as a function of the lowest \ion{H}{i} column density. The dot dashed curve represents the DLA comoving line density obtained by integrating $f(N_{\ion{H}{i}},X)$ over a range of column densities. However, an extrapolation of the DLA distribution by a single power law may not be the best to represent the frequency distribution of \ion{H}{i} column densities at $z\sim5$ all the way down to $N_{\ion{H}{i}}<10^{19}cm^{-2}$ due to the flattening of f(N$_{\ion{H}{i}}$,X) at the start of the optically thick regime at the Lyman limit \citep{crighton}. The $dn/dX$ of LLS from \citet{crighton} at $z\sim5$ is shown in Figure \ref{fig:dla and sub-dla} using a black star and it does not coincide with the extrapolated DLA distribution. Each of the horizontal lines in the figure corresponds to different LIAs comoving line density at $z>5.7$ from this work. It is evident that the minimum hydrogen column density required to reproduce the observed comoving line density decreases from stronger \ion{Mg}{ii} ($W>0.3$\AA) to \ion{O}{i} and then to \ion{C}{ii} and weak \ion{Mg}{ii} ($W<0.3$\AA). The $dn/dX$ values of \ion{C}{ii} and \ion{Mg}{ii} are higher than the previous surveys of \citetalias{cooper} and \citetalias{chen} due to increased sensitivity and completeness correction of our survey. 

If the LIAs follow the f(N$_{\ion{H}{i}}$,X) distribution of DLAs in \citet{noterdaeme2009}, then the minimum column density of hydrogen required to reproduce the observed number densities for \ion{O}{i}, log $N_{\ion{H}{I},\text{min}}/\text{cm}^2\ga19.25$; for \ion{C}{ii} and weak \ion{Mg}{ii}, log $N_{\ion{H}{I},\text{min}}/\text{cm}^2\ga18.85$ and for stronger \ion{Mg}{ii}, log $N_{\ion{H}{I},\text{min}}/\text{cm}^2\ga19.60$. \ion{O}{I} and stronger \ion{Mg}{ii} fall in the sub-DLA category (in agreement with the predictions of \citet{becker2011} for \ion{O}{i} and \citetalias{cooper} for stronger \ion{Mg}{II} where they used the $dn/dX$ from \citetalias{chen}). We present additional evidence for \ion{O}{I} tracing sub-DLAs, $dn/dX$ for \ion{O}{I} at $z>5.7$ is $0.24^{+0.16}_{-0.10}$ which is close to the combined number density of DLA and sub-DLA over $3<z<5$ \citep{omeara2007, prochaska, noterdaeme2009, crighton2015}. Weak \ion{Mg}{ii} and \ion{C}{ii} fall in the LLS category contrary to the prediction of \citetalias{cooper} where they used the $dn/dX$ value for \ion{C}{ii} from their work. \citetalias{cooper} predicted the high redshift \ion{C}{II} systems existing as DLAs and sub-DLAs. Dividing the \ion{Mg}{ii} sample based on their equivalent widths, the weak absorbers with $dn/dX=0.357$ falls within the column densities of LLS or super-LLS (sub-DLAs) systems probably tracing the infalling and co-rotating gas in the galaxy halos as predicted by \citet{kapzak2011,kapzak2012, nielsen2015}. The medium absorbers fall in the sub-DLA systems and for the strong absorbers, there are no detections at $z>5.7$, but by using the upper limit (see Table \ref{tab:4}), they correspond to the \ion{H}{i} column density of sub-DLAs. Thus, it is an additional evidence apart from the histogram in Figure \ref{fig:W norm hist}, for the incidence rates of \ion{Mg}{ii} being largely dominated by the weak absorbers at high z. As the equivalent width of the \ion{Mg}{ii} absorbers decreases, they tend to be increasingly ionized. This is also supported by the work of \citet{nielsen} where they showed an anti-correlation between the equivalent width of the \ion{Mg}{ii} systems and the impact parameter and thus, exposing the weaker absorbers more to the ionizing UV background. Moreover, \citet{stern} found that the density profile of the cool gas scales inversely with distance from the galaxy centre manifesting that the strong LIAs lie close to the galaxy itself. 

We also tried to obtain neutral hydrogen column densities of \ion{Mg}{ii} systems using the scaling relations in \citet{menard2009} and \citet{lan} in order to compare with the log $N_{\ion{H}{I},min}/\text{cm}^2$ values obtained from Figure \ref{fig:dla and sub-dla}. According to \citet{lan}, the scaling relation between \ion{Mg}{II} equivalent width and \ion{H}{I} column density is \begin{equation}
    N_\ion{H}{I}=A\bigg(\frac{W_{\lambda2796}}{1\text{\AA}}\bigg)^\alpha(1+z)^\beta
\end{equation} where $\alpha=1.69\pm0.13$, $\beta=1.88\pm0.29$ and $A=10^{18.96\pm0.10}\text{cm}^{-2}$. Applying this equation to our weak \ion{Mg}{ii} data, we obtain $\text{log}N_{\ion{H}{I},min}/\text{cm}^2 =19.22$ and for \ion{Mg}{II} with $W\ge0.3\text{\AA}$, $\text{log}N_{\ion{H}{I},min}/\text{cm}^2 =20.18$. These values are very close to the inferred neutral hydrogen column densities in Figure \ref{fig:dla and sub-dla}. Previously, using low redshift ($z<1.65$) \ion{Mg}{ii} absorbers with $W>0.3\text{\AA}$, \citet{menard2009} developed a scaling relation given as \begin{equation}
    \langle N_\ion{H}{I}\rangle_g(W_0)=C_g(W_0)^{\alpha_g}
\end{equation} where $C_g=(3.06\pm0.55)\times10^{29}\text{cm}^{-2}$ and $\alpha_g=1.73\pm0.26$. Applying this relation to weak and stronger absorbers provided us with \ion{H}{I} column densities for weak absorbers as $\text{log}\ N_{\ion{H}{I},min}/\text{cm}^2 =18.12$ and for stronger absorbers, $\text{log}\ N_{\ion{H}{I},min}/\text{cm}^2 =19.11$. These results are almost one order of magnitude lower than the values from Figure \ref{fig:dla and sub-dla} which might be because \citet{menard2009} based their analysis only on low redshift stronger absorbers. 

Having seen that the weak \ion{Mg}{ii} and \ion{C}{ii} have similar \ion{H}{I} column densities, it is important to understand more about their ionisation conditions by looking at their association with high ionisation absorbers such as \ion{C}{IV} and \ion{Si}{IV} (In this analysis, we have considered only those \ion{C}{II} and \ion{O}{I} absorbers that can be detected across the redshift range where the associated ions are detected). We found that 50\% (11/22) of \ion{C}{ii} systems have both \ion{C}{IV} and \ion{Si}{IV} across $4.317\le z \le 6.339$ and $4.907\le z \le 6.339$ respectively. The \ion{C}{II}-\ion{C}{IV} association is closer to weak \ion{Mg}{II}-\ion{C}{IV} (43\%) while the fraction of \ion{C}{II} with \ion{Si}{IV} is almost double the fraction of weak \ion{Mg}{II}-\ion{Si}{IV} (25\%) association. Similarly, we analysed the association of \ion{O}{i} with high ionisation species to compare with the ionisation conditions of stronger \ion{Mg}{II} with $W>0.3\text{\AA}$. 38\% (3/8) of \ion{O}{I} absorbers is associated with \ion{C}{IV} while 25\% (2/8) of \ion{O}{I} absorbers have \ion{Si}{IV} association. Therefore, fraction of \ion{O}{I} with \ion{C}{IV} is more than half of stronger \ion{Mg}{ii}-\ion{C}{IV} (59\%) systems while the fraction of \ion{O}{I} absorbers associated with \ion{Si}{IV} is almost similar to stronger \ion{Mg}{II} (29\%). Overall, weak \ion{Mg}{II} \& \ion{C}{ii} and stronger \ion{Mg}{II} \& \ion{O}{I} have similar ionisation conditions although both ions in each pair appearing in the same systems is very rare. Only 5 out of 22 \ion{C}{II} has weak \ion{Mg}{II} associated with them while 3 out of 10 \ion{O}{I} systems are detected with stronger \ion{Mg}{II} at a redshift range of $5.169<z<6.381$.

 As one would expect, the Universe was largely neutral right after the formation of the first stars and galaxies during the EoR. 
The LIAs, especially \ion{C}{ii} and \ion{O}{i} trace neutral regions of the CGM that are slightly ionized. Nevertheless, \ion{C}{ii} is also found in somewhat ionized gas, more so than \ion{O}{i}. The ionization potential of these absorbers is less than 13.6 eV and therefore, they are not self-shielded by hydrogen and appear as singly ionized species in an otherwise neutral medium \citep{becker2015}. An increasing trend in the comoving line density of the LIAs towards high redshift indicates that some changes are happening to the ionizing radiation at $z\gtrsim 5.7$. In the early Universe, the ionizing radiation is not strong enough to produce highly ionized absorbers in the metal poor CGM and as a result, there existed a combined effect of low metallicity and weak ionizing photons towards the tail end of EoR responsible for the observed trends in \ion{C}{ii} and \ion{O}{i}.

The rising trend in the comoving line density of \ion{C}{ii} in this work provides further evidence to the findings of \citetalias{cooper} in which the high redshift absorbers are found to be dominated by the low ionization species. Furthermore, \citet{Davies2023} also found that the comoving line density of \ion{C}{iv} declines with increasing redshift using E-XQR-30 sample concurring with the work of \citetalias{cooper}. The presence of a weaker or softer ionizing background can produce an increase in \ion{C}{II} and a decrease in \ion{C}{iv} content at high z. 
Furthermore, an upturn is observed for \ion{O}{i} absorbers at $z>5.7$ from the E-XQR-30 sample when combined with the results of \citetalias{becker2019} at $z<5.7$.
Due to the similar ionization potential of \ion{O}{i} as of neutral hydrogen, it traces gas that is largely neutral. Therefore, its decline at $z<5.7$, points to the decline of the neutral regions and the transition of metals in the galaxy halos to higher ionization states due to the ongoing reionization process at $z\sim6$. 

The upturn in the \ion{O}{i} comoving line density suggests a rapid change in the ionization state of the CGM which can be produced by an external ionizing source rather than the host galaxy. According to \citet{harikane2018,harikane2022}, no significant changes are reported in the galaxy properties at $2<z<7$ that would otherwise potentially create a rapid inside-out ionization of the CGM gas. Moreover, \citetalias{becker2019} discusses the possibility of simultaneous reionization of the CGM and IGM and also an alternative scenario where the CGM remains self-shielded for a while when the local IGM undergoes reionization. In the second scenario, the ionization of the CGM occurs towards the tail end of EoR when the mean free path of photons increases \citep{fan2006, becker2021, gnedin, gaikwad, Zhu2023}, exposing the region to photons from distant sources. The rapid change in the ionization state of the galaxy halos is further supported by the abrupt increase in the volume averaged neutral hydrogen fraction across $5.7<z<6.4$ (similar to the redshift range of \ion{O}{i} where an upturn is found) shown with a fully coupled radiative hydrodynamic simulation known as Cosmic Dawn III (CoDa III) by \citet{Lewis_2022}. Additional evidence for this rapid increase can be found in \citet{gaikwad}, where they used quasar absorption spectra from XSHOOTER and ESI and modelled the fluctuations in ionizing radiation field using the post-processing simulation code “EXtended reionization based on the Code for Ionization and Temperature Evolution” (\textsc{EX-CITE}). All these simulations and observations including this work, collectively point out that the ionizing radiation underwent a significant strengthening at $z<5.7$ and predicts a late end of EoR towards $z\sim5.3$. The rapid increase observed in the comoving line density of \ion{O}{i} across both $z>5.3$ and $z>5.7$ (see Figure \ref{fig:7} and Table \ref{tab:3}) from this work points to persisting fluctuations in neutral hydrogen fraction until $z\sim5.3$. A late end to the EoR has also been observed by the recent works of \citet{zhu} and \citet{bosman2022} using the quasar spectra from E-XQR-30. 

\begin{figure*}
     \centering
     \begin{subfigure}[b]{0.4\textwidth}
         \centering
         \includegraphics[width=\columnwidth]{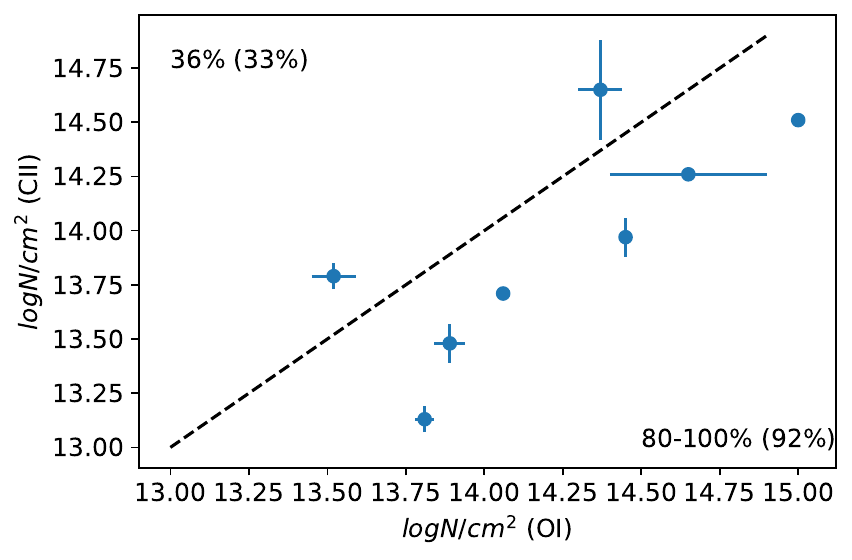}
         \caption{}
         \label{fig:CII/OI}
     \end{subfigure}
     \hfill
     \begin{subfigure}[b]{0.4\textwidth}
         \centering
         \includegraphics[width=\columnwidth]{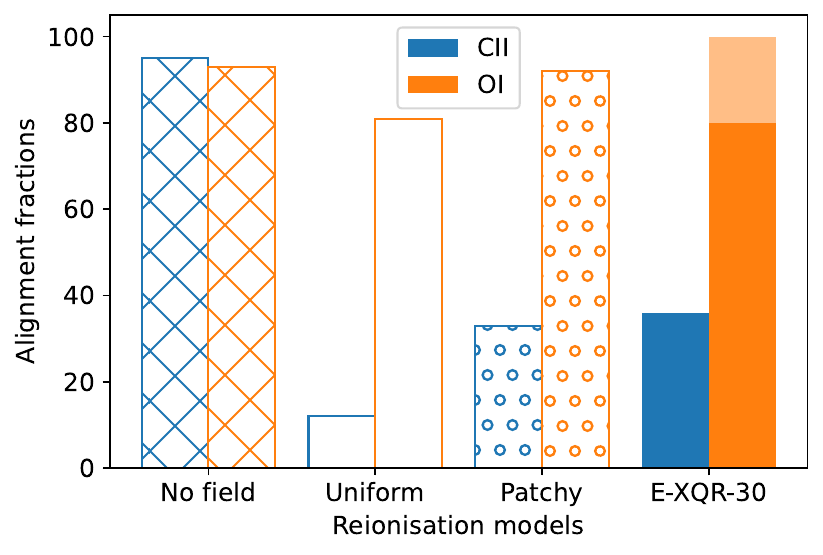}
         \caption{}
         \label{fig:alignment frac}
     \end{subfigure}
     \caption{The left panel represents the column density ratios of \ion{C}{ii} and \ion{O}{i} that are aligned to each other within a single system. The upper right corner in the figure shows the alignment fraction of \ion{C}{ii} with \ion{O}{i} (8/22) and the lower right shows the alignment fraction of \ion{O}{i} with \ion{C}{ii} (8-10/10). The values in brackets refer to the fractions from the simulations for a patchy reionisation model in \citet{oppenheimer2009}. The right panel shows the comparison of alignment fractions from different reionisation models in \citet{oppenheimer2009} with the observed  absorber ratios from E-XQR-30 and our results favour a patchy reionisation model. }
     \label{absorber alignment}
     \end{figure*}

\citet{oppenheimer2009} employed the column density ratios of aligned absorber as a statistical tool to compare the different ionisation models. Applying similar technique to our observations, we plot the column density ratios of \ion{C}{ii} and \ion{O}{i} that fall in the same system as defined by our survey in Figure \ref{fig:CII/OI} to compare it with the three models from \citet{oppenheimer2009}. 8/22 \ion{C}{II} absorbers are aligned with \ion{O}{i} while 8/10 \ion{O}{I} absorbers are associated with \ion{C}{ii}. The remaining 2 \ion{O}{i} absorbers are also aligned with \ion{C}{ii}, however, those two \ion{C}{II} absorbers are coincident with BAL regions of the quasar and therefore, are not included in the primary sample of the E-XQR-30 catalog. Thus, the alignment fraction of \ion{C}{II} with \ion{O}{I} is 36\% and \ion{O}{I} with \ion{C}{II} ranges from 80\% to 100\%. 
If the Universe was fully neutral at $z\sim6$ with no ionising background existing below the Lyman limit, we would expect a large fraction of \ion{C}{ii} absorbers to show associated \ion{O}{i} absorption but this is not the case in our work. We have only detected 22 \ion{C}{II} and 10 \ion{O}{i} intervening absorbers in the primary sample at $5.2<z<6.4$. In the second scenario, the Universe is assumed to have been fully ionised with a spatially uniform background everywhere and the third case is a patchy reionisation model. Figure \ref{fig:alignment frac} clearly demonstrates that the fraction of \ion{C}{ii} aligned with \ion{O}{i} and vice-versa from the simulation for the patchy reionisation model is in agreement with the observed alignment fractions. The hatched bins represent the alignment fractions of different reionisation models from \citet{oppenheimer2009} and the colour filled bins represent the observed ratios from this work. Therefore, our results favour a patchy reionization rather than a spatially homogeneous ionising background. Moreover, the non-detection of \ion{O}{i} in many sightlines where \ion{C}{ii} and other high ionisation species are detected, indicate that those sightlines probe ionised regions while others pass through metal enriched neutral gas close to the galaxies. A caveat to the reionisation models in \citet{oppenheimer2009} is that those models are not based on a radiative transfer simulation, so self-consistently accounting for self-shielding could change the alignment results. 

\begin{figure*}
     \centering
     \begin{subfigure}[b]{0.4\textwidth}
         \centering
         \includegraphics[width=\textwidth]{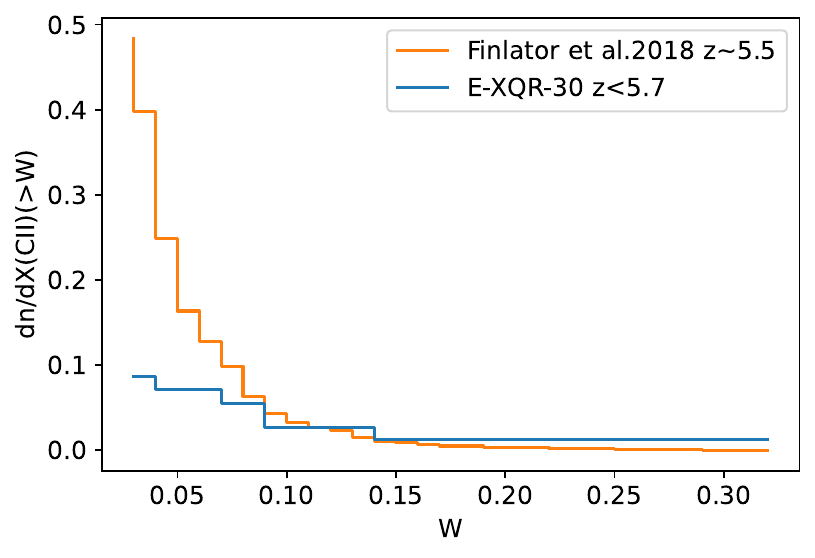}
         \caption{}
         \label{fig:W_dist_CII_low_z}
     \end{subfigure}
     \hfill
     \begin{subfigure}[b]{0.4\textwidth}
         \centering
         \includegraphics[width=\textwidth]{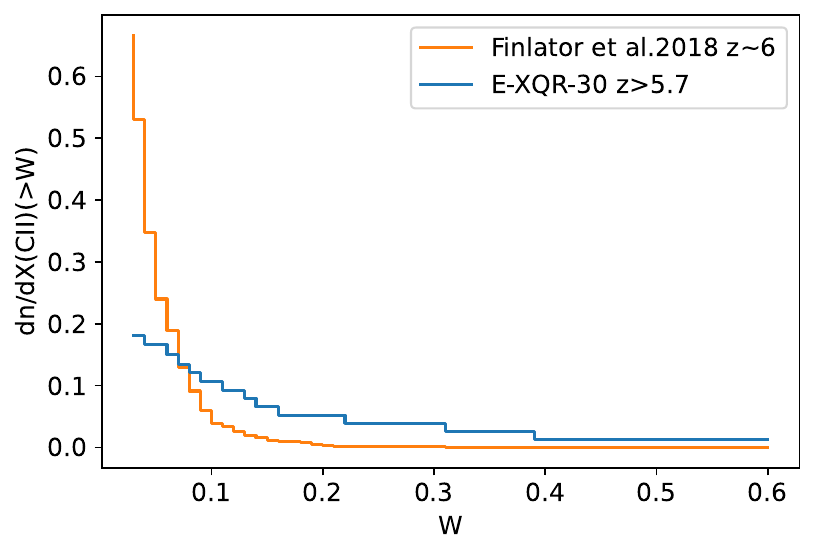}
         \caption{}
         \label{fig:W_dist_CII_high_z}
     \end{subfigure}
     \hfill
     \begin{subfigure}[b]{0.4\textwidth}
         \centering
         \includegraphics[width=\textwidth]{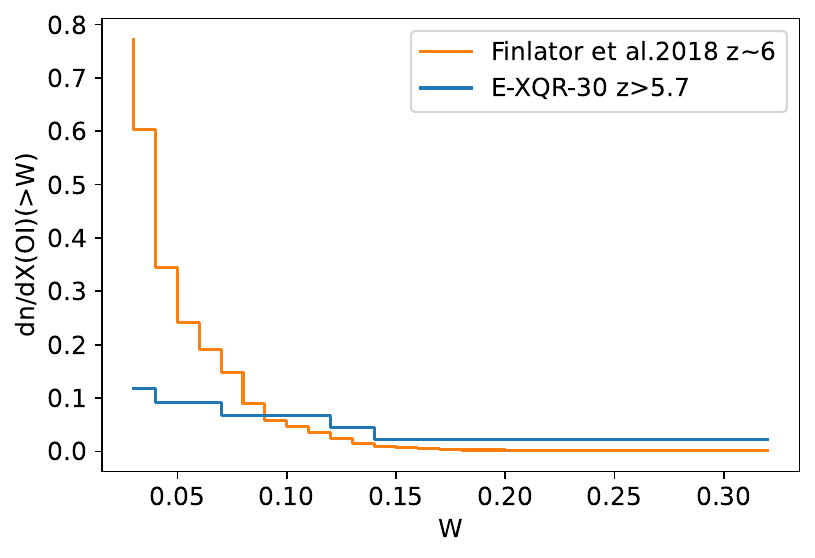}
         \caption{}
         \label{fig:W_dist_OI_low_z}
     \end{subfigure}
     \caption{The equivalent width distribution for low ionisation species for comparison with simulations from \citet{finlator2018}. The top panels show the distribution for \ion{C}{ii} and the bottom panel illustrates the distribution for \ion{O}{i}. The orange coloured histogram represents the model from \citet{finlator2018} while the blue histogram stands for the results from this work. It can be seen that both \ion{C}{ii} and \ion{O}{i} data have been divided into redshift intervals for appropriate comparison with the simulations at those redshifts. The simulations overpredict the weak absorbers while they agree with the observed distribution towards higher equivalent widths at $W>0.1\text{\AA}$. For \ion{O}{i}, comparison of the simulations with the lower redshift sample is not shown here due to very small number of absorbers at $z<5.7$. }
        \label{fig:W_dist_CII_OI}
\end{figure*}

Using the Technicolor Dawn simulations, \citet{finlator2018}, produced a spatially inhomogeneous reionisation model for the galaxy-driven reionisation scenario where they compare their simulated equivalent width distribution with the then existing observations for \ion{C}{ii} and \ion{O}{i} such as \citet{becker2011, bosman2017}. They found that the simulations overproduced weaker systems and underproduced strong systems. However, using data from E-XQR-30, we computed the $W$ distribution by calculating the comoving line densities for absorbers with different minimum equivalent widths as shown in Figure \ref{fig:W_dist_CII_OI}. The top panels show the distribution for \ion{C}{ii} equivalent widths and the bottom panel demonstrates the equivalent width distribution for \ion{O}{i}. The blue histograms indicate the observed distribution from this work while the orange histograms show the simulated distribution from \citet{finlator2018}. The observed equivalent widths for \ion{C}{II} and \ion{O}{I} have been divided at $z\sim5.7$ for comparison with the corresponding simulations. However, histograms comparing \ion{O}{I} distribution at $z<5.7$ is not included in the figure because there are not enough absorbers at this redshift range. Comparing with simulations from \citet{finlator2018}, the model overpredicts the weak absorbers but is in agreement towards higher equivalent widths at $W>0.1\text{\AA}$ for both \ion{C}{ii} and \ion{O}{i}. Therefore, the distribution from our work agrees better with the simulations when compared to the earlier works. The over prediction of weak absorbers in the simulation could be due to issues related to CGM metallicity, wind speed and UV background \citep{finlator2018}. 

\section{Conclusions}
\label{sec:Conclusions}

The evolution of low ionization absorbers, namely \ion{Mg}{ii}, \ion{C}{ii} and \ion{O}{i}, across redshift is studied using the E-XQR-30 metal absorber catalog prepared by \citet{xqr30catalog} from 42 high signal to noise ratio and intermediate resolution XSHOOTER quasar spectra at $z\sim6$. The catalog consists of a total of 778 systems including 260 \ion{Mg}{ii}, 22 \ion{C}{ii} and 10 \ion{O}{i} intervening absorbers separated from the background quasar by $>$10,000 \kms. The E-XQR-30 has  significantly increased the pathlength of $z\sim6$ \ion{C}{ii} and \ion{O}{i} by 50\% as well as the sample size of high redshift metal absorbers compared with previous large surveys of high redshift metal absorbers. For example, 131 weak \ion{Mg}{ii} ($W<0.3$\AA) absorbers at $z>2$ are detected for the first time indicating the improved sensitivity of E-XQR-30 over other high redshift metal absorber surveys. The sample completeness is generally high, reaching 50\% completeness at $W>0.03$\AA. 

\ion{Mg}{ii} absorbers, altogether, decline in comoving line density ($dn/dX$) towards high redshift with the $dn/dX$ of weak and medium absorbers remaining constant across redshift while the $dn/dX$ of strong absorbers decline with redshift. The cosmic mass density of \ion{Mg}{ii} ($\Omega_\ion{Mg}{ii}$) dominated by the strong \ion{Mg}{ii} absorbers follows the declining trend in the global star formation history across $2<z<5.5$. This is evidence that strong absorbers mimics the SFR evolution suggesting a connection between star formation and CGM enrichment rates. The weak \ion{Mg}{ii} sample in E-XQR-30 is of great significance as this is the first ever detection of a substantial size of population of these absorbers at $2<z<6$. For weak \ion{Mg}{II} systems to continue to trace sub-Lyman Limit System (LLS) environments to high redshift, their covering fraction must decline by factor of 3 as redshift 6 is approached. Their associations with \ion{Fe}{ii} and high ionization absorbers such as \ion{C}{iv} and \ion{Si}{iv} have also been studied respectively, in this work, providing insights into possible scenarios for the enrichment of these weak absorbers. The column density ratios of \ion{Fe}{ii} over weak \ion{Mg}{ii} indicates that they retained their enrichment mostly from core collapse supernovae of early stars particularly at $z>4.5$. \ion{C}{iv} and \ion{Si}{iv} associations indicate that these systems have high column densities of highly ionized absorbers and therefore, are exposed to ionizing radiation more than strong absorbers. The \ion{C}{ii} absorbers tend to increase with increasing redshift and the \ion{O}{i} absorbers show an upturn at $z>5.7$. The results obtained in this work is consistent with what has been observed in the previous high redshift surveys, in particular, \citetalias{chen} for \ion{Mg}{ii}, \citetalias{cooper} for \ion{C}{ii}, \citetalias{becker2019} for \ion{O}{i} and JWST results from \citet{Christensen2023} and emphasise the need for higher resolution spectroscopy to detect metal absorbers.

 The redshift evolution of metal absorbers provides information about the chemical enrichment of the Universe towards the tail end of EoR and the nature of ionizing radiation that ionized the neutral hydrogen in the CGM and IGM. The overall trend in the low ionization absorbers suggests a weaker UV ionizing radiation at $z>5$. As more galaxies formed, the metal content of the Universe increased through stellar nucleosynthetic processes and was ejected into the surrounding media. The results also favour the assumption of reionization of galaxy halos from external sources rather than in-situ sources. If the CGM is reionized by the host galaxies, then the ionization in CGM would tend to respond to the host galaxy properties in place of global evolution in the UV background. The decline in \ion{O}{i} at $z<5.3$ can also be related to the intensifying of the ionizing UV background towards $z<5.3$ and such a strengthening is expected only towards the end of reionization. Therefore, the rapid rise in the comoving line density in the low ionization absorber evolution also supports the recent studies proposing a late end of reionization, probably at $z\sim5.3$. Furthermore, our results provide evidence for this late reionization continuing to occur in metal-enriched and therefore biased regions of the Universe.

 The reionization of the Universe is still continuing to be an intriguing question for astronomers which created a major landmark in the transformation of the Universe from a homogeneous environment to a largely structured cosmos. The results from this work motivate future research on the relative role of chemical enrichment and nature of the UV background, both strength and hardness of the photons, during the early epochs which shaped the Universe as we see it today. This can be analysed by extending the sample size of \ion{C}{ii} absorbers that has a shorter pathlength with the help of a proxy and looking at the ratios of the low and high ionization states of carbon. The addition of more sightlines to deeper redshifts will also provide additional evidences to validate the current assumptions on the nature of reionization. Also, with the help of MUSE (Multi Unit Spectroscopic Explorer) \citep[for e.g.,][]{Diaz2021}, ALMA (Atacama Large Millimetre/submillimetre Array) \citep[for e.g.,][]{Kashino2023} and JWST (James Webb Space Telescope) \citep[for e.g.,][]{Bordoloi2023}, images and spectra of galaxies associated with the high redshift absorbers can be obtained using which the nature of the galaxies producing these absorbers in the early Universe can be identified. 

\section*{Acknowledgements}

The authors thank the referee for providing valuable feedback. This research was supported by the Australian Research Council Centre of Excellence for All Sky Astrophysics in 3 Dimensions (ASTRO 3D), through project number CE170100013. Based on observations collected at the European Organisation for Astronomical Research in the Southern Hemisphere under ESO Programme IDs 0100.A- 0625, 0101.B-0272, 0102.A-0154, 0102.A-0478, 084.A-0360(A), 084.A-0390(A), 084.A-0550(A), 085.A-0299(A), 086.A-0162(A), 086.A-0574(A), 087.A-0607(A), 088.A-0897(A), 091.C-0934(B), 096.A-0095(A), 096.A-0418(A), 097.B-1070(A), 098.B-0537, 098.B-0537(A), 1103.A-0817, 294.A-5031(B), 60.A-9024(A) and AMS acknowledges the XQR-30 collaboration involved in taking the observations and reducing the data. This work used NASA's Astrophysics Data System as well as \textsc{MATPLOTLIB} \citep{hunter}, \textsc{NUMPY} \citep{harris} and \textsc{SCIPY} \citep{scipy}. 

AMS acknowledges Asif Abbas for providing the results on cosmic mass density of \ion{Mg}{ii} at low redshift before publishing and the literature values for $dn/dX$ of \ion{Mg}{II}. AMS also acknowledges Kristian Finlator for providing comparison data from the simulations for the equivalent width distributions of \ion{O}{I} and \ion{C}{II}. 

For the purpose of open access, LCK has applied a Creative Commons Attribution (CC BY) licence to any Author Accepted Manuscript version arising from this submission. 

GK is partly supported by the Department of Atomic Energy (Government of India) research project with Project Identification Number RTI~4002, and by the Max Planck Society through a Max Planck Partner Group.

RAM acknowledges support from the Swiss National Science Foundation (SNSF) through project grant 200020\_207349.

SEIB is supported by the Deutsche Forschungsgemeinschaft (DFG) under Emmy Noether grant number BO 5771/1-1.

We also thank Nikole Nielsen, Glenn Kacprzak and Chris Blake for valuable discussions. 

\section*{Data Availability}


The metal absorber catalog used in this project and the Python script used to calculate the absorption path and redshift intervals are publicly available and can be downloaded at \url{https://github.com/XQR-30/Metal-catalogue/tree/main/AbsorptionPathTool}.



\bibliographystyle{mnras}
\bibliography{revised2} 




\appendix

\section{The redshift dependence of Mg {\scriptsize II} equivalent width distribution}
\label{MgII distribution with z}

The exponential distribution for \ion{Mg}{ii} equivalent widths $W>0.3$\AA\ has been found to steepen with redshift from the works of \citetalias{chen}. It is important to see whether this increase in slope with redshift will be observed when weak absorbers ($W<0.3$\AA) are also included in the sample. We attempt to fit for the equivalent widths of \ion{Mg}{ii} from E-XQR-30 with W>0.03\AA\ across two redshift ranges; $1.94<z<4.05$ and $4.05<z<6.38$ (the absorbers in the masked redshift intervals are not included) using the Schechter function given in equation \ref{eq:schechter}. Here, $W^*$ is kept constant at 1.44 (the best fitting $W^*$ for the total \ion{Mg}{ii} sample) to see whether the distributions steepen with redshift. 

\begin{figure}
    \centering
    \includegraphics[width=\columnwidth]{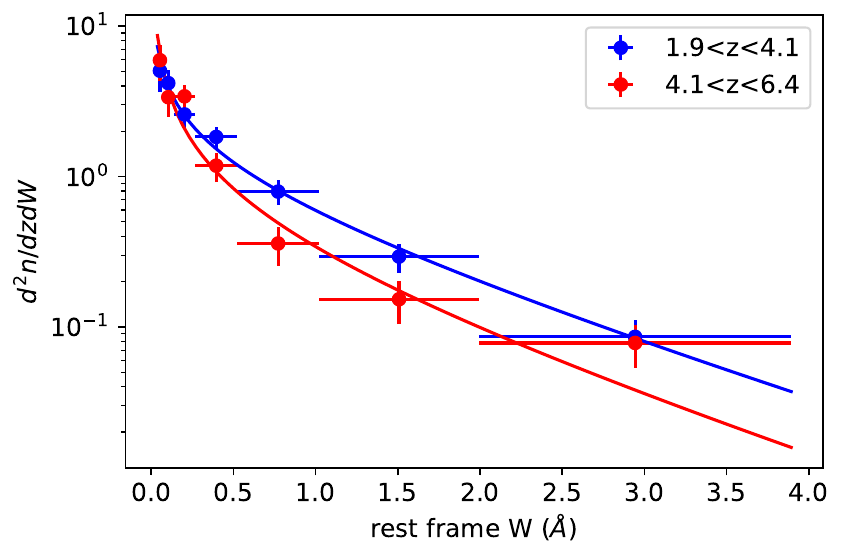}
    \caption{The equivalent width distribution fit at different redshift ranges using a Schechter function. The low redshift values and fit are shown in blue and the high redshift values and the fit are shown in red. The distribution is measured to steep in slope with increasing redshift.}
    \label{fig:W dist with z}
\end{figure}

The best-fit parameters are given in Table \ref{tab:best fit parameters}. \begin{table}
 \caption{The best fit parameters for the W distribution at different redshift ranges}
 \label{tab:best fit parameters}
 \begin{tabular}{|l|c|c|c|}
  \hline
  $z$ range & parameters & best fit value & \multicolumn{1}{|p{2cm}|}{\centering$1\sigma$ errors\\ (+,-)} \\
  \hline
  1.9-4.1 & $\Phi^*$ & 1.39 & 0.06,0.18  \\
   & $\alpha$ & -0.57 & 0.05,0.07\\
  \hline
  4.1-6.4 & $\Phi^*$ & 0.74 & 0.14,0.09  \\
   & $\alpha$ & -0.79 & 0.10,0.07\\
  \hline 
 \end{tabular}
\end{table}  The difference in parameters at different redshift intervals suggests that the equivalent distribution might not be the same at all redshifts. The fits obtained are shown in Figure \ref{fig:W dist with z}. The low redshift bin values and the corresponding fit are shown in blue while the high redshift bin values and their fit are represented in red colour. The power law slope $\alpha$ increases from -0.57 to -0.79 with redshift indicating a steepening of the weak absorber distribution as redshift increases. However, for the high redshift bin, the Schechter function fails to fit the values towards large equivalent widths where the distribution starts to flatten due to the small number of high redshift absorbers detected at these equivalent widths. 

\section{The comoving line density of O {\scriptsize I} using a different completeness correction}
\label{new OI dn/dX}

One of the reasons for the discrepancy in the $dn/dX$ values between this work and \citetalias{becker2019} can be associated to the completeness estimates in \citetalias{becker2019} where they expect some errors. To understand whether errors in the completeness calculations affect their results, they changed the minimum equivalent width of 25\% ($W>0.05$\AA) used in their analysis. \citetalias{becker2019} increased the completeness limit to 55\% ($W>1.0$\AA) which reduced the completeness corrections and still were able to recover the upturn in comoving line density at $z>5.7$, but at lower statistical significance. This is similar to what has been observed in the $dn/dX$ for \ion{O}{i} in this work which uses a similar completeness limit of 50\%. Therefore, their results seem to be driven by errors in the completeness estimates for smaller values of equivalent width. 

The other possible reason for the difference between $dn/dX$ values from our work and \citetalias{becker2019} might be due to the application of a fixed completeness correction as a function of equivalent width. There is a vague possibility of completeness varying with wavelength (redshift) even outside the masked regions of sky contamination due to changes in the S/N of the spectra at different wavelengths. Also, the completeness for \ion{O}{i} depends on the additional lines to be detected (e.g., \ion{C}{ii}, \ion{Si}{ii}). If one of these lines remains undetected due to the spectrum being noisy or contaminated at those wavelength regions, then \ion{O}{i} would not be identified. Therefore, the $dn/dX$ values from this work are recalculated after applying the completeness correction for \ion{O}{i} as a function of column density at different redshift intervals. The obtained values are very similar to the \citetalias{becker2019} values. However, the $dn/dX$ values obtained by applying completeness corrections as function of $W$ or log $N$ are consistent to each other within the error bars.

\begin{figure}
 \includegraphics[width=\columnwidth]{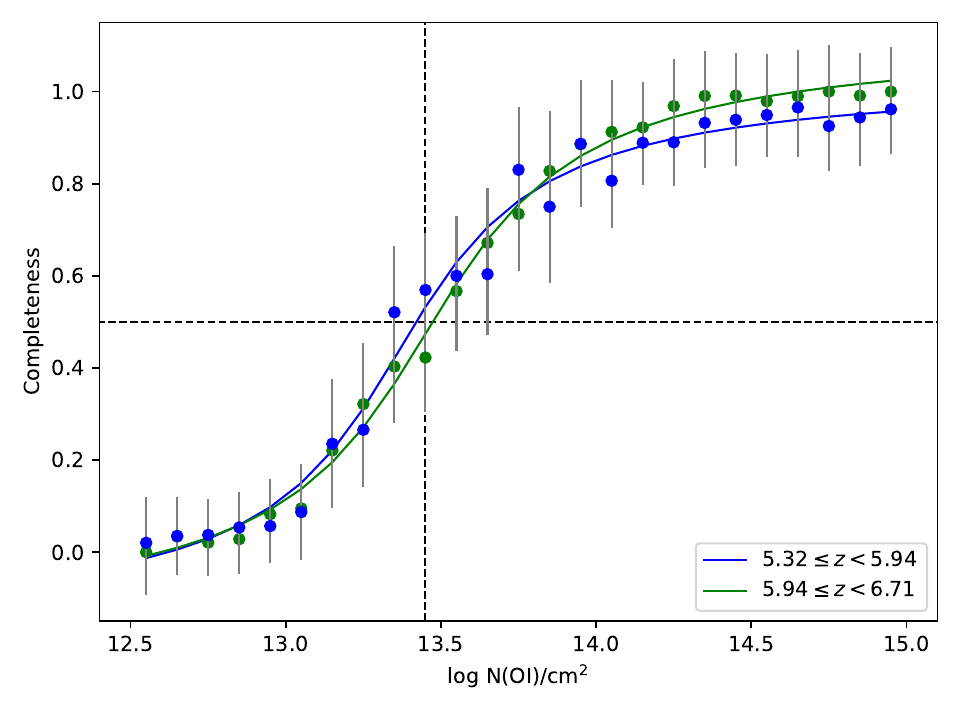}
 \caption{The figure shows the completeness correction for \ion{O}{i} as a function of $\text{log}\ N$.}
 \label{fig:OI completeness}
\end{figure}

\begin{figure}
    \centering
    \includegraphics[width=\columnwidth]{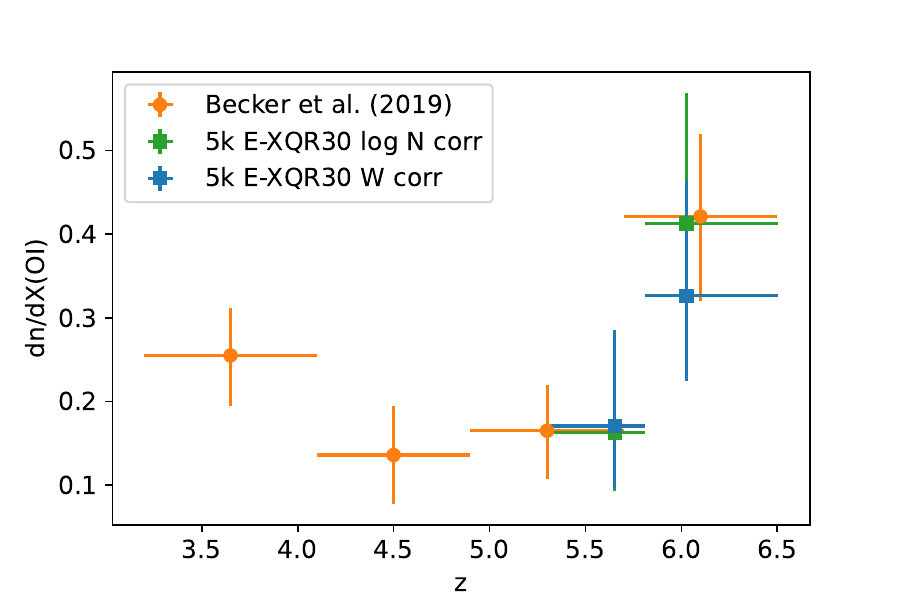}
    \caption{The comoving line density evolution of \ion{O}{i} across redshift using a proximity limit of 5000 \kms and completeness correction as a function of $\text{log}\ N$ is represented in green. The new $dn/dX$ values agree well with the \citetalias{becker2019} value at $z>5.7$. The $dn/dX$ values obtained after applying a completeness correction as a function of $W$ are shown in blue. However, both blue and green points at $z>5.7$ are consistent within the 1$\sigma$ error bars.}
    \label{fig:new OI dn/dX}
\end{figure}

The $dn/dX$ for \ion{O}{i} is calculated after correcting for completeness using the completeness correction as function of equivalent width. The obtained $dn/dX$ at $z>5.7$ did not match with the results in \citetalias{becker2019}. In the efforts to figure out the reason behind the discrepancy, the completeness correction for \ion{O}{i} is computed as a function of log $N$ at two redshift intervals; $5.32<z<5.94$ and $5.94<z<6.71$. The function used to correct for completeness of the sample is as follows. \begin{equation}
    \text{Completeness(log}N)=S_y(\text{arctan}(S_xW+T_x)+T_y)
\end{equation}The parameters obtained are given in Table \ref{tab:OI completeness}. \begin{table}
 \caption{The parameters used in the completeness correction for \ion{O}{i} as a function of $\text{log}\ N$}
 \label{tab:OI completeness}
 \begin{tabular}{|l|c|c|c|c|}
  \hline
  $z$ range & $T_x$ & $S_x$ & $T_y$ & $S_y$ \\
  \hline
  5.32-5.94 & -39.76 & 2.98 & 1.15 & 0.38 \\
  5.94-6.71 & -36.43 & 2.71 & 1.16 & 0.41 \\
  \hline
 \end{tabular}
\end{table} Figure \ref{fig:OI completeness} shows the new completeness corrections for \ion{O}{i} as a function of $\text{log}\ N$.  The \ion{O}{i} absorber data from the catalog is processed in a similar manner as outlined in Section \ref{subsec: The low-ionization absorbers: dn/dX}. In this case, the 50\% completeness limit for \ion{O}{i} is $\text{log}\ N/cm^2$=13.45. The $dn/dX$ values obtained after applying the new completeness correction are shown in Figure \ref{fig:new OI dn/dX}.  The green squares indicate the values obtained when a completeness correction as a function of column density is applied and blue squares represent the values obtained using the completeness correction as a function of equivalent width. Using the new completeness corrections, the $dn/dX$ values from both works agree to each other in both redshift intervals. However, both values calculated using different completeness corrections are in agreement within the $1\sigma$ confidence limits showing that both corrections produce consistent results.


\bsp	
\label{lastpage}
\end{document}